\documentclass[preprint,12pt,authoryear]{elsarticle}

\usepackage{amssymb}
\usepackage{amsmath}

\usepackage{xcolor}

\usepackage{bm}
\newcommand{\bdt}{\bm{\delta t}}
\newcommand{\bth}{\bm{\vartheta}}
\newcommand{\bL}{\bm{\Lambda}}
\newcommand{\tL}{\tilde{\bm{\Lambda}}}

\usepackage{lineno}

\usepackage{tikz}
\usetikzlibrary{shapes}
\usetikzlibrary{shapes.geometric}
\usetikzlibrary{arrows}
\usetikzlibrary{arrows.meta}
\usetikzlibrary{positioning}
\usetikzlibrary{calc}
\usetikzlibrary{backgrounds}

\usepackage[margin=2cm]{geometry}

\usepackage{array}
\usepackage{booktabs}

\usepackage{listings}

\usepackage{hyperref}
\usepackage[utf8]{inputenc}
\usepackage[T1]{fontenc}
\usepackage{textcomp}  
\usepackage{aas_macros} % add this

\usepackage{subfig}
\usepackage{xcolor}
\newcommand{\orcid}[1]{%
  \ifx&#1&%
  \else%
    \textcolor{green!50!black}{\footnotesize\href{https://orcid.org/#1}{[ORCID: #1]}}%
  \fi%
}

\journal{Astronomy and Computing}

\begin{document}

\begin{frontmatter}



\title{Addressing prior dependence in hierarchical Bayesian modeling for PTA data analysis II: Noise and SGWB inference through parameter decorrelation} %% Article title
%speeding up 


\author[iasf]{Eleonora Villa\corref{cor1}}
\ead{eleonora.villa@inaf.it}
\cortext[cor1]{Corresponding author.}

\author[koexai]{Luigi D'Amico}
\author[koexai]{Aldo Barca}
\author[koexai]{Fatima Modica Bittordo}
\author[koexai]{Francesco Al\`i}
\author[oas]{Massimo Meneghetti}
\author[koexai]{Luca Naso}

\affiliation[iasf]{organization={INAF – Istituto di Astrofisica Spaziale e Fisica Cosmica di Milano (IASF-MI)},
            addressline={via Alfonso Corti 12}, 
            city={Milano},
            postcode={20133}, 
            country={Italy}}

\affiliation[koexai]{organization={Koexai srl},
            addressline={Via Jose Maria Escriva, 6}, 
            city={Catania},
            postcode={95125}, 
            country={Italy}}

\affiliation[oas]{organization={Osservatorio di Astrofisica e Scienza dello Spazio di Bologna (INAF-OAS)},
            addressline={via Gobetti 93/3}, 
            city={Bologna},
            postcode={40129}, 
            country={Italy}}

%\affiliation[iasf]{organization={
%INAF – Istituto di Astrofisica Spaziale e Fisica cosmica di Milano (IASF-MI)},%Department and Organization
%            addressline={via Alfonso Corti 12}, 
%            city={Milano},
%            postcode={20133}, 
%            country={Italy}}

%% Abstract
\begin{abstract}
Pulsar Timing Arrays (PTA) provide a powerful framework to measure low-frequency gravitational waves, but accuracy and robustness of the results are challenged by complex noise processes that must be accurately modeled. 
Standard PTA analyses assign fixed uniform noise priors to each pulsar, an approach that can introduce systematic biases when combining the array.
To overcome this limitation, we adopt a hierarchical Bayesian modeling strategy in which noise priors are parametrized by higher-level hyperparameters. 
To mitigate the sensitivity of the inferred parameters to the choice of 
noise hyperprior, we introduce a reparametrization of the hierarchical 
model based on the orthogonal projection of hyperparameters onto the 
physical parameter subspace. The transformation is implemented through 
Normalizing Flows (NFs), which provide an invertible, tractable 
representation and preserve shrinkage and inter-pulsar information pooling 
in the reparametrized model.
We also employ \texttt{i-nessai}, a flow-guided nested sampler, to efficiently explore the resulting higher-dimensional parameter space. 
We apply our method to a minimal 3-pulsar case study, performing a simultaneous inference of noise and stochastic gravitational wave background (SGWB) parameters. 
Despite the limited dataset, the results consistently show that the reparametrized hierarchical treatment constrains the noise parameters more tightly and partially alleviates the red-noise–SGWB degeneracy, while the orthogonal reparametrization further enhances parameter independence without affecting the correlations intrinsic to the power-law modeling of the physical processes involved.
\end{abstract}

\begin{graphicalabstract}
\end{graphicalabstract}

\begin{highlights}
\item Research highlight 1 : Implementation of Hierarchical Bayesian modeling and integration of flow-guided nested sampling into Pulsar Timing Array Bayesian inference pipelines
\item Research highlight 2 : Decorrelation in the parameter space implemented via Normalizing Flows to address prior dependence
\end{highlights}

%% Keywords
\begin{keyword}
Hierarchical Bayesian modeling \sep Pulsar Timing Array \sep Nested sampling \sep Normalizing Flows \sep Decorrelation in the parameter space 



\end{keyword}

\end{frontmatter}

%\linenumbers

\section{Introduction}
Understanding physical phenomena through parameter inference from observational data is a core task in modern astronomy. To account for poorly constrained effects, unmodeled physics, noises, and/or some systematics, it is standard practice to introduce so-called nuisance parameters that capture unknown or subdominant contributions. Although not of direct scientific interest, nuisance parameters are often correlated with the target physical parameters, and this gives rise to degeneracies in the posteriors. As a result, the inferred values of the parameters of interest may be biased or misrepresented, depending on the modeling chosen for the nuisance contributions. This issue is particularly critical when the data lack the constraining power to disentangle nuisance effects from the parameters of interest.

This problem plays a particularly important role in the context of Pulsar Timing Arrays (PTA) analyses. PTA experiments measure the times of arrival (TOAs) of radio pulses emitted by highly stable millisecond pulsars. These measurements are modeled using a deterministic timing model that accounts for effects such as the pulsar spin evolution, astrometric parameters, and proper motion. Any deviation between the observed TOAs and the timing model predictions — referred to as timing residuals — is attributed to a combination of noise processes and potential gravitational wave signals. Recently, PTA experiments across the world have reported evidence for a stochastic gravitational wave background (SGWB), marking a major milestone in low-frequency gravitational wave astronomy, see \citet{Agazie2023evidence}, \citet{EPTACollaboration2023}, \citet{Xu2023}, and \citet{Reardon2023}. However, extracting meaningful information from the SGWB signal critically depends on accurately modeling the various noise contributions in the data, making this an especially delicate task. PTA analyses rely on modeling multiple sources of noise, such as intrinsic red noise and dispersion measure (DM) variations, which are both modeled as power-law signals and described by two parameters each, the $\log_{10}$ of the amplitude and the spectral index $\gamma$, for each pulsar. The same kind of parameters describe the SGWB signal too, as it is modeled as a single power law common process across all the pulsars in the array. In a single-pulsar analysis, each pulsar represents a single realization of data with respect to the noise stochastic processes. The crucial point here is that it could happen that a certain type of noise is not significantly detectable in the data of a single pulsar whereas it becomes detectable in the data of the full array: this means that the entire array of pulsars represents a different realization of data with respect to the same noise parameters. By contrast, since the SGWB is modeled as a single common signal shared across all pulsars, the full array provides a single realization with respect to the SGWB parameters, unlike the case for noise parameters. The above considerations imply that using the results of single-pulsar analyses — such as assuming the absence of a given noise process — as priors in a full-array analysis can be problematic. It is not only an example of the so-called circular use of data but it also applies a prior derived from the wrong statistical realization. The discrepancy between the statistical description of the noise processes in the single pulsars and their statistical description in the entire array may accumulate across pulsars and lead to biased inference, e.g. it may introduce systematic errors in the inferred values of $\log_{10}A$  and $\gamma$ for the power law model of the SGWB. This is especially crucial for the pulsar red noise which is highly correlated with the SGWB, both being low-frequency stochastic processes modeled in the same way. The solution of this problem is then clear: the distribution of the noise parameters should be parametrized for each pulsar and inferred simultaneously with that of the two SGWB parameters in the analysis of the full array. This means that we have to choose a prior for all the parameters to infer, including noise priors. To account for the uncertainty in the noise distributions we parametrize noises prior by introducing an additional hierachical prior encoded in additional parameters, called hyperparameters. They describe the distribution of noise properties across the array and act as nuisance parameters in the inference problem. We remark that the hierarchical Bayesian modeling becomes necessary in PTA analyses, as it allows for the correct inference of both the astrophysical signal of interest, such as the SGWB, and the population-level noise properties encoded in the hyperpriors. All the above considerations were recently pointed out, although with slightly different approaches, in \citet{ensemble}, \citet{rutger-hier}, and \citet{rutger-av}. For a discussion about previous literature implicitly showing this problem, referred to as \textit{prior misspecification}, also see \citet{ensemble}.

At this point, a well-known aspect of Bayesian modeling becomes relevant: posterior distributions can be significantly influenced by the choice of priors. This problem, known as \textit{prior dependence} or prior-driven projection effects, affects every level of hierarchical modeling.
Various strategies have been explored to address this issue in many different contexts. In general, one approach is to impose physically motivated priors, for instance informed by simulations or empirical models. Another possibility is to employ priors considered "non-informative", such as uniform or Jeffreys priors. Yet these choices come with their own limitations: uniform priors may implicitly encode arbitrary assumptions e.g. on prior bounds, and Jeffreys priors, although invariant under reparametrization, are often impractical in complex analyses. While uniform priors are of standard use in pulsar astronomy, in \citet{gibbs} is proposed the use of Jeffreys-like prior for all the red processes (red noise and SGWB) in the full array analysis. At the level of hierarchical Bayesian modeling, the \textit{prior dependence} can be formulated as follows: the inference of the parameters of physical interest - the noises and the SGWB parameters in the case of PTA - is sensitive to the specific prior that we choose - the hyperprior for the noises in the case of PTA. This is recovered in the work by \citet{ensemble}, where the authors resolve \textit{prior misspecification} by introducing hyperparameters for red and DM noises.
Then, they employ a novel numerical marginalization technique over hyperparameters for the inference of physical parameters of the red and DM variation noises but still their results are sensitive to the choice of the higher-level prior of the noises.

In this work, we use the hierarchical Bayesian description but we pursue a different strategy. 
Rather than proposing specific hyperpriors or marginalizing over hyperparameters, our approach aims to reduce prior sensitivity by working at the level of the hierarchical model itself. We develop a reparametrization designed to weaken the coupling between physical noise parameters and nuisance hyperparameters, which aims to make the inference less dependent on the specific choice of noise hyperpriors, i.e to mitigate \textit{prior dependence}. Our methodology builds on the principle of parameter orthogonalization, first established in a formal statistical framework by \citet{cox-reid-1987}, who demonstrated that orthogonal parameterizations can achieve approximate conditional inference and reduce parameter correlations. This foundational concept was subsequently extended by \cite{tibshirani-wasserman-1994}, who explored various theoretical and practical aspects of statistical model reparameterization, and by \cite{christensen-etal-2006}, who showed that orthogonal transformations significantly enhance MCMC sampling efficiency. More recently, \cite{Paradiso_2025} applied a non-linear orthogonalization technique using Generalized Additive Models (GAMs) to the Effective Field Theory of large-scale structure in cosmology. They showed that decorrelating cosmological and nuisance parameters makes the posterior less sensitive to the choice of nuisance priors. Their results agree with the theoretical predictions of \cite{cox-reid-1987}, who established a comprehensive framework demonstrating that orthogonal parameterizations in hierarchical models naturally lead to more robust inference with decreased prior sensitivity. Our contribution consists of applying this approach to the hierarchical structure of the PTA noise model. The idea is conceptually quite simple: we orthogonalize the hyperparameter vector by removing its projection onto the subspace spanned by the physical noise parameters. The resulting orthogonal complement vector constitutes our new decorrelated hyperparameter vector, which is, by construction, decorrelated from the physical parameters. To implement the orthogonal transformation we employ a 2-steps Normalizing Flows algorithm that learns how to pass from one parameter set to the other. We then systematically evaluate how this reparameterization affects the posterior distributions of the noise parameters.
For Bayesian inference, we use \texttt{i-nessai}, a flow-guided nested sampler presented in \citet{Williams_2023}, which updates the previous version of the algorithm, see \citet{Williams_2021} and \citet{nessai}. \texttt{i-nessai} leverages Normalizing Flows to navigate complex, high-dimensional parameter spaces efficiently, making it particularly advantageous for computationally intensive likelihood evaluations characteristic of PTA analyses. The method provides substantial computational acceleration compared to the standard parallel-tempered Markov chain Monte Carlo (\texttt{PTMCMC}) sampler traditionally used in PTA studies, \cite{nessai-spoke3}.

The structure of the paper is as follows. In Section~\ref{sec:PTA-setup} we introduce the setup PTA data analysis and outline the role of hierarchical Bayesian modeling. In section~\ref{sec:orto} we summarize the key steps of the reparametrization procedure and its implementation. A full and detailed description is provided in our companion paper \citet{HBM-method}.
In Section~\ref{sec:results} we discuss the results obtained from a minimal 3-pulsar case study, illustrating the main effects of the hierarchical modeling and reparametrization on noise and SGWB inference. Finally, Section~\ref{sec:conclusions} provides our conclusions and perspectives for future applications.
We complement the paper with three appendices: \ref{app:projection_algebra} summarizes the orthogonal projection algebra, \ref{app:prior_sensitivity} discusses prior sensitivity in the reparameterized hierarchical model, and \ref{app:state_plot} presents the \texttt{i-nessai} diagnostic plots for the hierarchical runs.

\section{PTA data analysis set up}\label{sec:PTA-setup}
In PTA experiments the primary observables are the times of arrival (TOAs) of pulses from an ensemble of millisecond pulsars. These TOAs are compared to a deterministic timing model that accounts for pulsar spin-down, astrometric, and (where relevant) binary parameters. The differences between the observed TOAs and the predictions of the timing model define the timing residuals, which contain both astrophysical signals and various noise contributions. The residuals can be decomposed into several components: (i) the timing model itself, which is linearized around best-fit parameters; (ii) intrinsic red noise, a correlated stochastic process associated with each pulsar individually; (iii) a stochastic gravitational wave background, modeled as a red process correlated across all pulsars according to the Hellings–Downs correlation; (iv) DM variations noise, due to the frequency-dependent propagation of the radio signal through the ionized interstellar medium; and (v) white noise, typically dominated by radiometer noise and possible short-term pulse-phase jitter, both uncorrelated between TOAs. For a detailed description of PTA data model and likelihood, we refer the reader to Chapter 7 of \citet{taylor}. 

In our hierarchical description of the noise, we focus on intrinsic red noise and DM variations noise. White noise, being uncorrelated with both red processes and the SGWB, is fixed (we set EFAC to unity) and not included in the reparametrization scheme. 
Our analysis is based on synthetic pulsar timing data generated from the European Pulsar Timing Array (EPTA) DR2new parameter files \citep{DR2new}. 
We simulate timing residuals for an array of millisecond pulsars with realistic observational properties: a cadence of 5 days, observation frequencies randomly drawn from $\{500, 900, 1400\}$~MHz, timing uncertainties of $1~\mu$s, and a baseline spanning from MJD~52000 to~59000 (approximately 19 years). 
Using the \texttt{libstempo} framework, we inject multiple stochastic processes to construct a realistic test scenario, including intrinsic red noise, dispersion measure variations, and a stochastic gravitational-wave background. 
All processes are modeled as Gaussian processes with power-law power spectral densities of the form
\[
P(f) = \frac{A^2}{12\pi^2}\,\left(\frac{f}{f_{\rm yr}}\right)^{-\gamma} f^{-3},
\]
where $A$ and $\gamma$ denote the amplitude and spectral index, respectively. Each process is represented using 30 Fourier components over the frequency range $[10^{-9}, 10^{-5}]$~Hz. For DM variations, the additional chromatic dependence $f^{-2}$ of the timing residuals is accounted for through the Fourier design matrix, rather than explicitly included in the power spectral density. The injected parameters have values: $
A_{\rm RN} = 7\times10^{-14}, \gamma_{\rm RN} = 3;
A_{\rm DM} = 5\times10^{-14},\gamma_{\rm DM} = 2.3;
A_{\rm GWB} = 2\times10^{-15}, \gamma_{\rm GWB} = 13/3.$

This formulation is consistent across all red processes and ensures a uniform parametrization of the intrinsic, chromatic, and common signals. 
The complete probabilistic model is implemented through \texttt{Enterprise}, which constructs the total covariance matrix by combining the individual Gaussian-process components and analytically marginalizing over the linear timing model parameters. 
For each pulsar, intrinsic red noise and DM variations are described by independent Gaussian processes with parameters $(\log_{10} A_{\rm RN}, \gamma_{\rm RN})$ and $(\log_{10} A_{\rm DM}, \gamma_{\rm DM})$, respectively. 
Uniform priors are assigned as
$\log_{10} A_{\rm RN}, \log_{10} A_{\rm DM} \sim \mathcal{U}(-16, -12)$ and $\gamma_{\rm RN}, \gamma_{\rm DM} \sim \mathcal{U}(0, 7)$.
The SGWB is modeled as a common Gaussian process shared by all pulsars, with spatial correlations described by the Hellings--Downs overlap reduction function. Its parameters $(\log_{10} A_{\rm GWB}, \gamma_{\rm GWB})$ follow the same uniform prior ranges as the intrinsic processes. The likelihood for PTA residuals is modeled as a multivariate Gaussian distribution with a covariance matrix that incorporates the correlations induced by the various processes. Standard PTA analyses marginalize analytically the linear timing model parameters, yielding a reduced likelihood that depends only on the stochastic processes implemented in a single object in \texttt{Enterprise}.

%%%%%%%%%%%%%%%%%%%%%%%%%%%%%%%%%%%%%%%%%%%%%%%%%%%%%%%%%%%%%%%%%

\section{Parameter decorrelation in hierarchical Bayesian modeling}\label{sec:orto}

\subsection{Methodology}\label{subsec:method}
Here we highlight the essential aspects of the reparametrization procedure, while the complete methodological description can be found in our companion paper \citet{HBM-method}.
Let us start by making the idea of parameter orthogonalization more formal. We consider some physical parameters $\bth$ whose prior distribution $\pi(\bth|\bL)$ is parametrized by the hyperparameters $\bL$ which in turn are distributed according to their hyperprior $\pi'(\bL)$, in general different from the distribution $\pi$. In our herarchical framework the total number of physical noise parameters is $n = n_p \times 4$, where $\bth = \{\gamma_{\rm RN}, \gamma_{\rm DM}, \log_{10}A_{\rm RN}, \log_{10}A_{\rm DM}\}$ and $p$ is the number of pulsars in the dataset. The hyperpriors can be chosen as Gaussian or uniform distributions, with 2 hyperparameters each - mean and standard deviation for the Gaussian prior or lower and upper bounds for the uniform prior -, leading to $m = 8$ hyperparameters $\bL$ in total. The full parameter space has dimension $n+2+m$, where the two parameters of the SGWB, $\gamma_{\rm GWB}$ and $\log_{10}A_{\rm GWB}$ common to all pulsars, are added to the $n$-dimensioanl set of the physical parameters $\bth$. Note, however, that the SGWB parameters are not involved in the hierarchical structure nor in the reparametrization procedure. The joint posterior of all the two-levels parameters is given by, see \citet{ThraneTalbot2019} and \citet{ensemble}
\begin{equation} \label{post}
\mathcal{P}(\bth,\bL|\bdt) = \frac{\mathcal{L}(\bdt|\bth)\pi(\bth|\bL)\pi'(\bL)}{\mathcal{Z}}\,.
\end{equation}
In the above equation $\mathcal{L}(\bdt|\bth)$ is the likelihood and depends on the physical parameters only and $\mathcal{Z}$ is the Bayesian evidence. The hierarchical structure is fully encoded in the parametrized conditional prior term $\pi(\bth|\bL)$ giving the distribution of $\bth$ conditional to that of the hyperparameters $\bL$, $\pi'(\tL)$. 

The basic idea for the decorrelation is simple: given the hyperparameter vector $\bL$ and the subspace generated by the physical noise parameters $\bth$, we obtain the decorrelated hyperparameters $\tL$ by computing the orthogonal complement of $\bL$ with respect to the $\bth$-subspace, effectively removing any linear dependence between the two parameter sets. The required transformation is given by
\begin{equation}\label{projection}
   {\tL}={\bL}-P_{\bth}\bL=(I-P_{\bth})\bL\,, 
\end{equation}
where
\begin{equation}
    P_{\bth}\equiv\bth(\bth^T\bth)^{-1}\bth^T
\end{equation}
is the projector on the subspace spanned by $\bth$. Geometrically, this means removing from $\bL$ the component lying along the direction of $\bth$. As a result, the transformed hyperparameters $\tL$ satisfy by construction the orthogonality condition $\bth^T \tL=0$, whose derivation is given in \ref{app:projection_algebra}.
The initial hierarchical structure encodes how physical parameters depend on potentially correlated hyperparameters, along with the hyperparameters own distribution. Our goal is to construct an equivalent representation by using the new $\tL$, defined to be orthogonal to the physical parameter directions, along with the corresponding distribution of these transformed hyperparameters. That is, we want to construct the transformation
\begin{equation}\label{transformation}
    \pi(\bth|\bL)\pi'(\bL) \quad\longrightarrow\quad\pi(\bth|\tilde\bL)\tilde\pi'(\tilde\bL)\,.
\end{equation}
Since the conditional prior $\pi(\bth|\bL)$ encodes the statistical relationship between physical parameters and hyperparameters in the hierarchical model, when Bayesian inference is performed in the reparametrized space, the posterior naturally inherits the decorrelated structure from the transformed conditional prior $\pi(\bth|\tL)$. This transformation should not be understood as removing the hierarchical dependence between physical parameters and hyperparameters. Rather, it defines a reparametrized hierarchical representation where transformed hyperparameters are designed to be less coupled to the physical parameters, while population-level information is still encoded in the learned conditional prior $\pi(\bth|\tilde{\bL})$ and in the transformed hyperprior $\tilde{\pi}'(\tilde{\bL})$. This is the sense in which our proposed method addresses prior dependence, i.e. the dependence of the posterior of the physical parameters on the functional form of the hyperprior: it is designed to mitigate nuisance-prior projection effects and, at the same time, improve the geometry explored by the sampler. The impact of our approach on prior sensitivity is discussed in detail in~\ref{app:prior_sensitivity}.

The issue with this simple procedure is that the orthogonal projection in Equation~\ref{projection} does not admit an inverse. As a consequence, the variable change in Equation~\ref{transformation} is ill-defined. We address this issue by directly modeling both $\pi(\tilde\bL)$ and $\pi(\bth|\tilde\bL)$ with machine learning generative algorithms that accurately approximate complex or even singular distributions into simpler ones through sequences of invertible mappings. They are called \textit{Normalizing} Flows exactly because of their ability to regularize irregular distributions; for reviews and applications in Bayesian inference see \citet{papaspiliopoulos-etal-2007}, \citet{rezende2015variational}, and \citet{Kobyzev_2021}. Here we also make use of a special subclass of NFs, especially designed for conditional probability distributions, \citet{trippe2018conditional}.
Our procedure is as follows. We sample from the prior $\pi(\bL)$, then propagate through the hierarchy to obtain $\bth$, and finally compute $\tilde\bL$ via the projection:
\begin{enumerate}
    \item draw $\bL_i\sim\pi(\bL)$ for $i=1,\ldots,N_{\text{samples}}$;
    \item draw $\bth_i\sim\pi(\bth|\bL)$ for $i=1,\ldots,N_{\text{samples}}$;
    \item compute $\tilde\bL_i=(I-P_{\bth_i})\bL_i$
    \item check the orthogonality condition $\bth_i^T \tL_i=0$ for the samples.
\end{enumerate}
It is important to remark here that the drawing in steps 1 and 2 is performed from the prior distributions of $\bL$ and $\bth$, rather than from their posteriors. This choice avoids any form of circular analysis, ensuring that the Normalizing Flows learn the mapping between hyperparameters and physical parameters without being biased by the specific data realization or by the posterior structure of a given dataset\footnote{The approach we use in this work is conceptually very close to the one of \citet{Paradiso_2025} but differs in three aspects: (i) we use Normalizing Flows instead of GAMs for integration with our nested sampling framework and exact Jacobian computation; (ii) we apply the reparametrization within a hierarchical structure and our hyperparameters are in the conditional prior $\pi(\bm{\theta}|\bm{\Lambda})$ and in the hyperprior $\pi(\bm{\Lambda})$ (iii) we do not sample from the posterior to train the NFs.}.

At this point, two complementary NFs learn the distributions needed for Equation~\ref{transformation} directly from the triples $(\bth_i,\bL_i,\tilde\bL_i)$. This dual construction provides the necessary ingredients for Equation~\ref{transformation}: by exploiting NFs, we approximate the mapping of Equation~\ref{projection} and obtain a regularized inverse, all while remaining consistent with the orthogonalization scheme and the hierarchical relation between priors and hyperpriors.
We stress that the use of Normalizing Flows in our framework is the key computational advantage: although the projection $\bL \to \tilde{\bL}$ does not admit an analytic inverse, the learned flow-based representation is invertible by construction and provides a tractable exact Jacobian, which carries the complete hierarchical structure. It is precisely through this property that the population structure is fully retained in the reparametrized model: population information continues to propagate to the lower-level noise parameters through the learned conditional prior, so that shrinkage and information pooling across pulsars are preserved.

\subsection{Implementation for PTA data analysis}
In this section we provide a schematic description of the implementation of our hierarchical framework for PTA data analysis, as pictured in Figure~\ref{fig:hier_diagram}. A comprehensive and technical discussion can be found in our companion paper~\citet{HBM-method}.

\smallskip
\noindent
\textbf{\textit{Normalizing Flows.}}
The orthogonal transformation described in Section~\ref{sec:orto} is implemented through a two-step NFs architecture that learns the mapping between the correlated and decorrelated representations of the hierarchical model. The flows learn to implement the transformation defined in Equation~\ref{transformation} where analytic inversion is impossible. The priors in the reparametrized space are defined implicitly by the trained flows, but this implicit definition arises from - and is constrained by -learning a distribution that explicitly enforces the orthogonality constraint $\bm{\theta}^T \tilde{\bm{\Lambda}} = 0$ at the training stage.

\smallskip
\noindent
The first component, the \textbf{\textit{push-forward}} flow (PF-NF), learns the distribution $\pi'(\tilde\bL)$ of the decorrelated hyperparameters, preserving orthogonality. It is implemented as a Masked Autoregressive Flow (MAF) \citep{papamakarios2017masked}, composed of three transformation blocks with masked affine autoregressive layers of 32 hidden units each. This design ensures exact invertibility with tractable Jacobian computation in $\mathcal{O}(m)$ time, enabling both efficient sampling and accurate likelihood evaluation.

\smallskip
\noindent
The second component, the \textbf{\textit{pull-back}} conditional flow (PB-CNF), learns the conditional distribution $\pi(\bth|\tilde\bL)$ of the physical parameters given the decorrelated hyperparameters, preserving the hierarchical structure. It is implemented as a conditional MAF \citep{trippe2018conditional}, where the context $\tilde{\bm{\Lambda}}$ is injected at each transformation layer. 
The network consists of three conditional masked affine transformations with shared weights across layers, allowing the model to capture the hierarchical dependence between the two parameter spaces while preserving orthogonality by construction.

\smallskip
\noindent
Both flows are trained jointly on a dataset of $N_{\text{samples}} = 20000$ realizations drawn from the priors of $(\tilde{\bm{\Lambda}}, \bm{\vartheta})$. Training is performed with a mini-batch size of 256, introducing stochasticity that helps avoid poor local minima and enhances generalization. 
A 90/10 split between training and validation sets is adopted, and an early-stopping criterion selects the model corresponding to the minimum validation loss. 
The optimization objective is the standard negative log-likelihood used in NF training: the PF-NF maximizes the likelihood of the decorrelated hyperparameters, while the PB-CNF maximizes the conditional likelihood of the physical parameters.

\smallskip
\noindent
\textbf{\textit{Implementation for PTA data analysis.}}
The reparametrized hierarchical Bayesian framework is implemented through the joint use of \texttt{Enterprise} and \texttt{i-nessai}: \texttt{Enterprise} serves as a likelihood engine, always evaluated exclusively for the physical parameters without altering the internal structure of the \texttt{PTA} class, and in accordance with the hierarchical posterior structure of Equation~\ref{post}. All prior and hyperprior specifications, both in their original and reparametrized forms, are defined and managed within \texttt{i-nessai}. This split implementation allows \texttt{Enterprise} to focus solely on the physical model likelihood, while \texttt{i-nessai} handles all aspects of hierarchical sampling and reparametrization. 
The resulting interface provides a modular and extensible framework for hierarchical posterior inference in PTA analyses, readily adaptable to larger or more complex datasets without modifying the internal \texttt{Enterprise} architecture.

\begin{figure}[htbp]

\centering

\begin{tikzpicture}[
    node distance=2.2cm,
    databox/.style={rectangle, draw, thick, fill=purple!10, minimum width=2cm, minimum height=0.9cm, align=center},
    methodbox/.style={rectangle, draw, thick, fill=green!10, minimum width=2cm, minimum height=0.9cm, align=center},
    likebox/.style={rectangle, draw, thick, fill=yellow!10, minimum width=3.2cm, minimum height=1cm, align=center},
    neuralbox/.style={rectangle, draw, thick, fill=orange!10, minimum width=3.4cm, minimum height=1.1cm, align=center},
    arrow/.style={->, thick, >=stealth},
    deflink/.style={->, semithick, dashed, >=stealth, gray},
    title/.style={font=\bfseries}
]

% Title

\node[title] at (-6, 10.5) {Architecture of i-nessai hierarchical models};

% Nodes

\node[neuralbox] (reparam) at (-6,9) {
    \textbf{Input Data}\\
    Noise Priors $\pi(\bm{\vartheta}\!\mid\!\bm{\Lambda})$ and   Hyperpriors $\pi'(\bm{\Lambda})$ \\
    or\\
    Reparametrizer (NFs): PB--CNF $\pi(\bm{\vartheta}\!\mid\!\tilde{\bm{\Lambda}})$ and PF--NF $\pi(\tilde{\bm{\Lambda}})$
};

\node[likebox] (prior) at (-8, 6.5) {\textbf{log\_prior}\\
  $\log \pi(\bm{\vartheta}\!\mid\!\bm{\Lambda})+\log \pi'(\bm{\Lambda})$\\
  or\\
  $ \log \pi(\bm{\vartheta}\!\mid\!\tilde{\bm{\Lambda}})+\log \pi'(\tilde{\bm{\Lambda}})$
};

\node[likebox] (params) at (-3, 6.5) {\textbf{Parameter Space}\\ 
$\bm{x}=(\bm{\vartheta}, \bm{\Lambda})$\\
or\\
$\bm{x}=(\bm{\vartheta}, \tilde{\bm{\Lambda}})$
};

\node[methodbox] (nessai) at (-10, 4) {\textbf{Sampler}\\i-nessai Nested Sampling\\Flow-based proposals};

\node[methodbox] (likelihood) at (-5, 4) {\textbf{log\_likelihood}\\
$\mathcal{L}(d\!\mid\!\bm{\vartheta})$};

\node[methodbox] (pta) at (-1.5, 4) {\textbf{Enterprise}\\PTA Model};

\node[databox] (output) at (-9, 1.5) {\textbf{Posterior Samples}\\
$P(\bm{\vartheta}, \bm{\Lambda} \mid d)$\\
or\\
$P(\bm{\vartheta}, \tilde{\bm{\Lambda}} \mid d)$
};

% Connections

\draw[arrow]   (reparam) -- (prior);
\draw[arrow] (params) -- (prior);
\draw[arrow] (prior) -- (nessai);
\draw[arrow] (params) -- (likelihood);
\draw[arrow] (likelihood) -- (nessai);
\draw[arrow] (pta) -- (likelihood);
\draw[arrow] (nessai) -- (output);

\end{tikzpicture}

\caption{Architecture of the \texttt{i-nessai} hierarchical model, shown in both the non-reparametrized (hyperparameters $\bL$) and reparametrized (hyperparameters $\tilde\bL$) forms. This implementation handles standard hierarchical Bayesian inference in the two situations: the first with explicit uniform conditional priors $\pi(\bm{\vartheta}|\bm{\Lambda})$ and Gaussian or uniform hyperpriors $\pi'(\bm{\Lambda})$; the second with a pair of Normalizing Flows: the push-formward PF-NF encoding $\pi'(\tilde\bL)$ and the pull-back PB-CNF encoding $\pi(\bm{\vartheta}|\tilde\bL)$. The likelihood is evaluated via \texttt{Enterprise} and priors and likelihood feed the sampler \texttt{i-nessai} in parallel. The output are the posterior samples.}

\label{fig:hier_diagram}
\end{figure}

\section{Discussion of the results}\label{sec:results}
In this Section we present the results of our hierarchical Bayesian analysis applied to the simulated PTA dataset introduced in Section~\ref{sec:PTA-setup}. 
We applied the proposed decorrelation scheme within the hierarchical framework to a three-pulsar configuration, performing a simultaneous inference of noise and SGWB parameters. 
Although such a small array is not sufficient to draw exhaustive quantitative conclusions, it provides a clear and controlled case study that allows us to isolate and highlight the main effects of the hierarchical modeling and reparametrization. 
Hyperparameters were introduced only for the noise components, namely for the intrinsic red noise and the DM variations, while the stochastic gravitational wave background parameters are left at the standard single-level description. 
We focus here on the representative pulsar J1744$-$1134. Analogous results are obtained for the other two pulsars in the array.

As a starting point, we consider the reference runs with fixed, non-hierarchical priors on the noise parameters, 
i.e.\ flat priors $\pi(\bth)$ within the bounds $[-16, -12]$ for $\log_{10}A$ and $[0, 7]$ for $\gamma$, as specified in Section~\ref{sec:PTA-setup}. 
This baseline configuration corresponds to the standard PTA inference approach, where all noise and signal parameters are treated independently, without introducing hyperpriors. The second introduces hierarchical priors $\pi(\bth|\bL)$ on the red-noise and DM variation parameters, with Gaussian hyperprior $\pi(\bL)$ across the array. The means and standard deviations of the Gaussian $\pi(\bL)$ are taken from the inference reported in Table 2 of \citet{ensemble}, ensuring consistency with population-level noise statistics derived from the EPTA dataset. The third configuration applies the orthogonal reparameterization $\pi(\bth|\tilde\bL)$ with hyperprior $\pi(\tilde\bL)$ , where hyperparameters are decorrelated from physical parameters through the projection implemented with Normalizing Flows, as described in Section~\ref{sec:orto}.

\begin{figure}[htbp]
  \centering
  \subfloat[
 Independence score metric values introduced by the hierarchical treatment of the noise before decorrelation. All noise parameters dispaly a mild interplay while the SGWB results to be almost uncorrelated.
  ]{
    \includegraphics[width=0.45\textwidth]{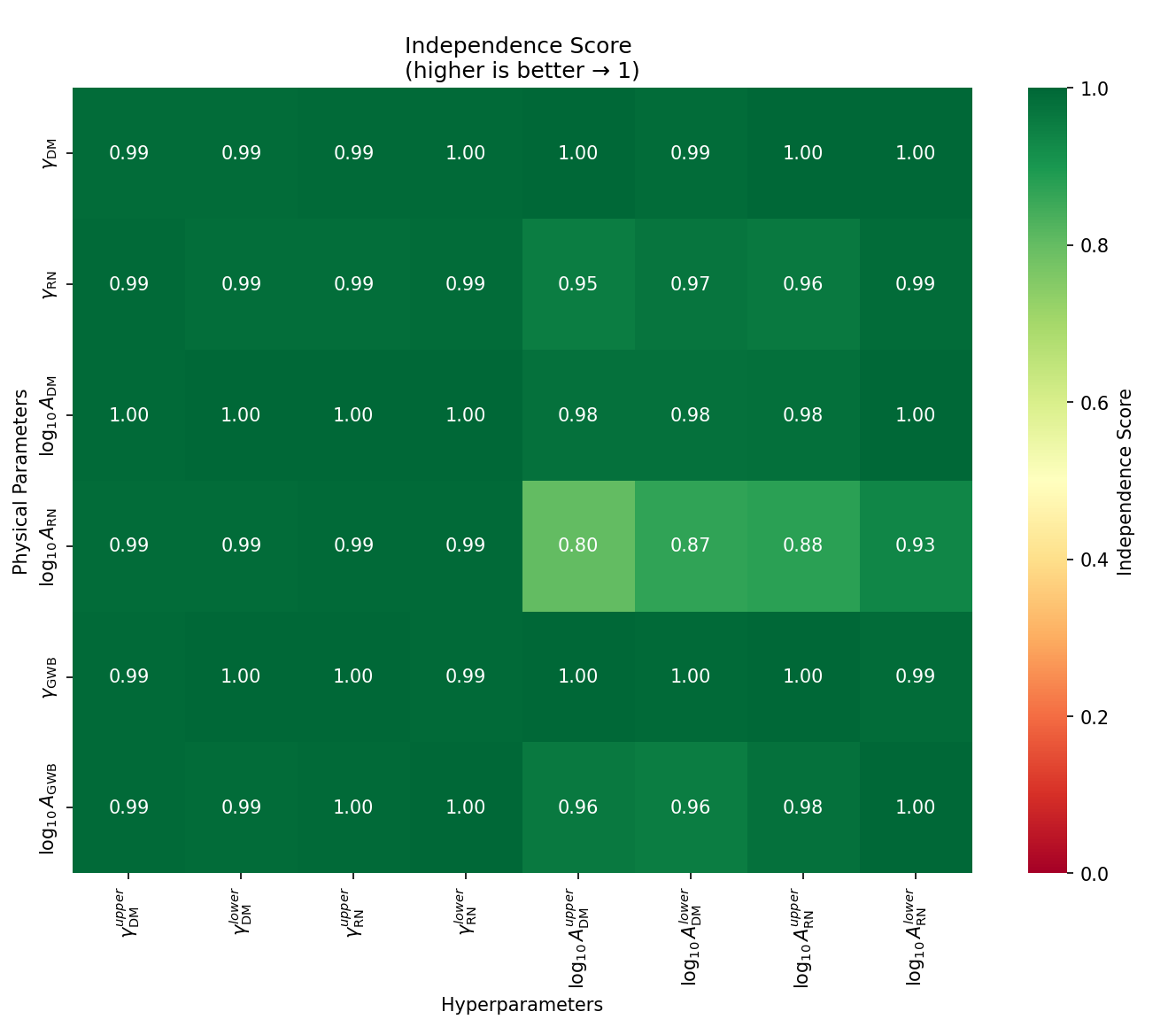} \label{fig:I-original}
  }
  \hfill
  \subfloat[
  %short: 
  Independence score metric values after hyperparameters decorrelation. All physical parameters are now effectively independent of their corresponding hyperparameters, 
  except for a mild residual correlation in $\log_{10}A_{\rm{RN}}$.
  ]{
    \includegraphics[width=0.45\textwidth]{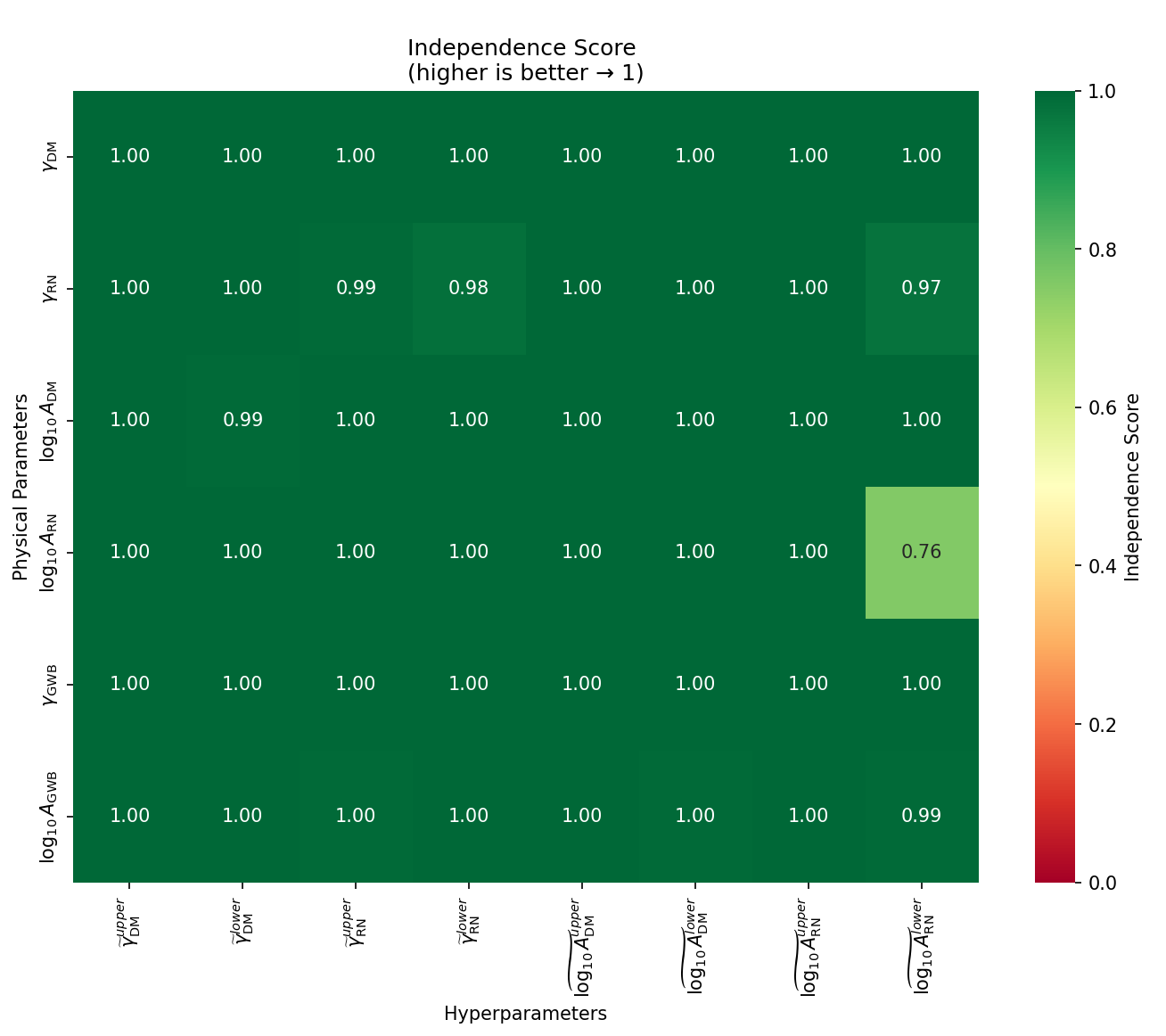}\label{fig:I-rep}
  }
    \caption{
    Independence score metric $\mathcal{I}(\vartheta_i, \Lambda_j)$ between physical noise parameters and hyperparameters for pulsar J1744$-$1134. 
    }  \label{fig:corner&indep-rep}
\end{figure}

To quantify the effect of the decorrelation procedure on the parameter--hyperparameter dependencies, we employ a detailed parameter–hyperparameter correlation analysis based on the independence score metric $\mathcal{I}(\vartheta_i, \Lambda_j)$ defined as 
\begin{equation} \label{eq:iscore}
    \mathcal{I}(\vartheta_i, \Lambda_j)=\frac{\mathbb{E}[\text{Var}(\vartheta_i|\Lambda_j)]}{\text{Var}(\vartheta_i)}
\end{equation}
and ranging from 0 to 1, with values approaching 1 indicating that $\vartheta_i$ is largely independent of $\Lambda_j$. 
This metric quantifies the fraction of variance in the physical parameter that is not explained by the hyperparameter, thus measuring the degree of statistical independence. Figure~\ref{fig:corner&indep-rep} shows the independence score computed for all parameter--hyperparameter pairs in both the standard hierarchical framework (panel~\ref{fig:I-original}) and after orthogonal reparametrization (panel~\ref{fig:I-rep}). 
In the original hierarchical configuration with $\pi(\bth|\bL)$, the noise parameters display mild correlations with their corresponding hyperparameters, as expected from the hierarchical coupling encoded in the conditional priors. 
After applying the orthogonal projection to obtain $\pi(\bth|\tilde\bL)$, all physical parameters become effectively independent of the transformed hyperparameters, with independence scores approaching unity. The only exception is $\log_{10}A_{\rm{RN}}$, which retains a mild residual correlation. This feature is already present in the original parameterization and is consistent with the intrinsic difficulty of constraining red noise amplitude parameter in small arrays. 
Notably, the SGWB parameters are essentially uncorrelated with the noise hyperparameters in both configurations, confirming that the hierarchical treatment of the noise sector does not introduce artificial dependencies on the common signal in this 3-pulsar setup. We conclude this discussion of the independence score metric with a 
methodological remark, since its interpretation is central to the 
compatibility of our reparametrization with the hierarchical nature of 
the model. The metric in Eq.~\eqref{eq:iscore} is a second-order, 
variance-based diagnostic: independence scores approaching unity indicate 
that the transformed hyperparameters $\tL$ carry essentially no variance-level information about the physical parameters $\bth$, but do not imply full 
statistical independence between the two sets. This is consistent with 
the sample-level orthogonality condition $\bth^T \tL = \mathbf{0}$ 
enforced by the projection (see~\ref{app:projection_algebra}), which is itself a statement 
about inner products rather than about statistical 
independence. A residual non-linear dependence of $\bth$ on $\tL$ 
survives the projection and is captured by the pull-back conditional flow PB-CNF representing $\pi(\bth|\tL)$: this encodes shrinkage and inter-pulsar information pooling in 
the reparametrized hierarchical model. The near-unity values in 
Figure~\ref{fig:corner&indep-rep} should therefore be read as evidence that 
the orthogonal projection has successfully removed the variance-level coupling between physical parameters and hyperparameters, 
while the hierarchical information flow between the two levels remains 
encoded in the non-linear structure of the learned conditional prior. For a 
detailed discussion of this interplay we refer to~\ref{app:prior_sensitivity}.

\begin{figure}[htbp]
\centering
\includegraphics[width=0.8\textwidth]{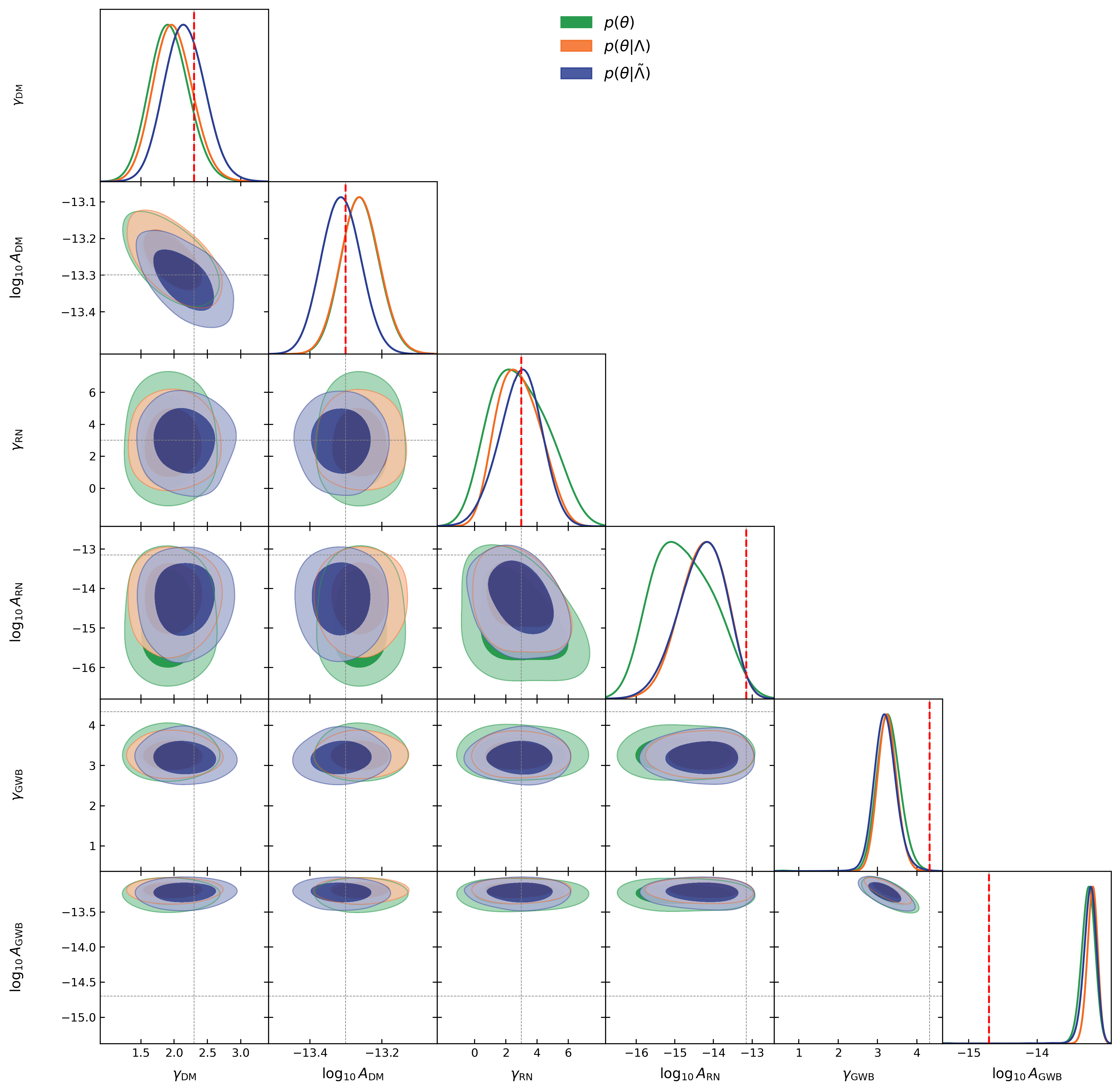}
\caption{Corner plot showing the 2D-joint posterior distributions of the noise and SGWB parameters for pulsar J1744$-$1134 under three different prior configurations. 
Green contours correspond to fixed uniform priors $\pi(\bth)$ without hierarchical structure, representing the standard PTA analysis approach. 
Orange contours show results with hierarchical priors $\pi(\bth|\bL)$ 
Blue contours display the posteriors after orthogonal reparametrization $\pi(\bth|\tilde\bL)$, where hyperparameters are decorrelated from physical parameters. 
The diagonal panels show the corresponding 1D marginal distributions. 
The hierarchical treatment and the decorrelation progressively tightens the red-noise constraints and reveals the characteristic $\log_{10}A$--$\gamma$ anticorrelation, while the SGWB parameters remain largely unchanged across all configurations in this minimal 3-pulsar setup.}\label{fig:corner}
\end{figure}

Figure~\ref{fig:corner} displays the 2D-joint posterior distributions for all noise and SGWB parameters across the three inference configurations. 
A clear feature visible across all panels is the characteristic anticorrelation between the spectral index and the logarithmic amplitude typical of power-law processes, see e.g.\ \citet{Lentati2016, taylor}. 
This negative correlation arises from the degeneracy in the power-law spectrum $P(f) \propto A^{2} f^{-\gamma}$: 
for a given dataset, a steeper spectral index (larger $\gamma$) can be compensated by a higher amplitude $A$, and vice versa. 
The effect is particularly evident for the SGWB, whose amplitude and spectral index show elongated, anti-correlated posterior contours. 
The same qualitative behavior is observed for the DM variation noise parameters $(\log_{10}A_{\rm{DM}}, \gamma_{\rm{DM}})$, 
although it is mitigated by the additional frequency dependence $f^{-2}$ characteristic of chromatic processes: 
since DM variations imprint a dispersive delay that scales with the inverse squared observing frequency, 
multi-frequency data provide an extra constraint that partially breaks the degeneracy between amplitude and spectral index.

Crucially, both the hierarchical treatment and the orthogonal reparametrization preserve this intrinsic $\log_{10}A$--$\gamma$ anticorrelation structure. 
This represents an important validation of our approach: the hierarchical framework and the decorrelation procedure do not disrupt the genuine correlations that arise from the power-law modeling of the underlying stochastic processes. 
In other words, the reparametrization acts only on the hierarchical layer, leaving intact the lower-level structure.

Beyond this anticorrelation feature, the corner plot reveals differences in the constraining power across different processes. 
In the baseline configuration with fixed uniform priors $\pi(\bth)$ (green contours), 
the intrinsic red noise is practically unconstrained, 
while the SGWB shows the opposite behavior, with sharply peaked posteriors for both $\log_{10}A$ and $\gamma$. 
This contrast directly reflects the different constraining power of the dataset on individual versus common processes. 
Since both the red noise and the SGWB contribute power on overlapping low-frequency ranges, 
the sampler tends to allocate the common signal either to the SGWB or to the individual red-noise terms 
depending on their prior bounds, number of Fourier components, correlation pattern, and on the number of pulsars in the array. 
With only three pulsars, the dataset provides limited leverage to disentangle these two sources of correlated noise, 
and the sampler naturally favors the SGWB component, whose spatial Hellings–Downs correlation pattern provides a more distinctive signature. 
With the common uniform prior range $\log_{10}A \in [-16,-12]$, the inferred SGWB amplitude is therefore strongly peaked toward the upper edge of the prior, 
reflecting the fact that part of the low-frequency power is absorbed by the red noise, while the remaining contribution is ascribed to the SGWB. 
This behavior, often referred to as the \textit{red noise–SGWB degeneracy} 
\citep[e.g.][]{vanHaasteren2014, EPTACollaboration2023}, 
motivates the introduction of a hierarchical description of the noise model, which allows us to capture the variability of the noise parameters and to mitigate prior-driven projection effects.

The introduction of the hierarchical structure with $\pi(\bth|\bL)$ (orange contours) significantly tightens the red-noise constraints: 
the marginal posteriors for both $\gamma_{\rm{RN}}$ and $\log_{10}A_{\rm{RN}}$ become noticeably more concentrated, 
and the characteristic $\log_{10}A$--$\gamma$ anticorrelation begins to emerge clearly. 
The orthogonal reparametrization with $\pi(\bth|\tilde\bL) $(blue contours) further enhances this effect, 
yielding even tighter constraints and more pronounced correlation structure for the red-noise sector. 
In contrast, the SGWB parameters remain essentially unchanged across all three configurations. 
This insensitivity reflects the fact that, in this minimal 3-pulsar setup, 
the limited size of the array prevents the hierarchical treatment from effectively breaking the red-noise--SGWB degeneracy at the level of the common signal.

\begin{figure}[htbp]
\centering
\includegraphics[width=0.9\textwidth]{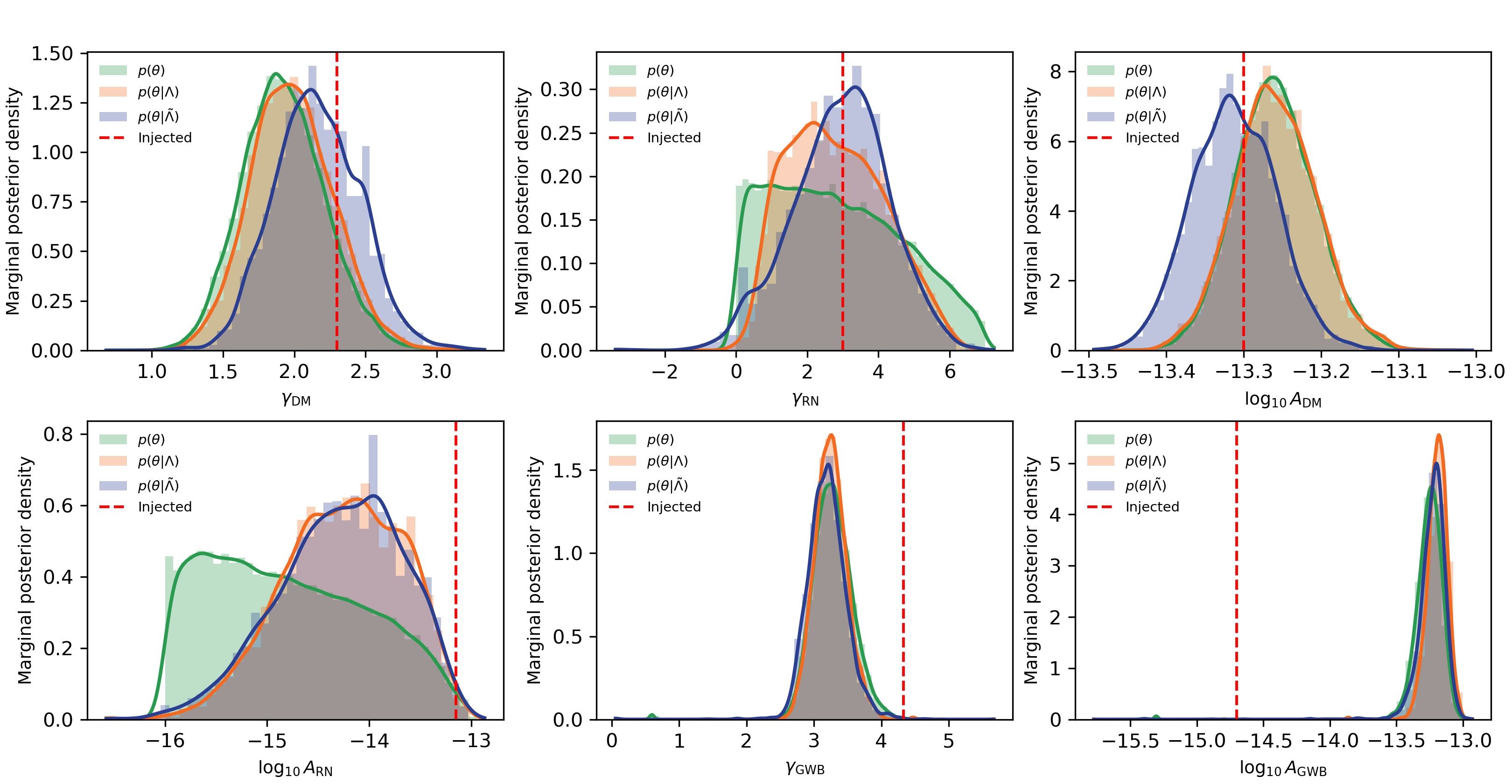}
\caption{Marginal posterior distributions for the noise and SGWB parameters of pulsar J1744$-$1134 comparing the three inference frameworks. 
Green curves show results with fixed uniform priors $\pi(\bth)$, orange curves with hierarchical priors $\pi(\bth|\bL)$, and blue curves after orthogonal reparametrization $\pi(\bth|\tilde\bL)$. 
Red dashed vertical lines mark the injected true values. 
The red-noise spectral index $\gamma_{\rm{RN}}$ and amplitude $\log_{10}A_{\rm{RN}}$ become progressively more constrained through the hierarchical treatment and decorrelation, recovering better agreement with the injected values. 
The DM variation posteriors show comparable widths in the three approaches but shift progressively toward the injected values through the hierarchical treatment and decorrelation. SGWB parameters show minimal variation across the three approaches, with the SGWB amplitude remaining biased toward the prior upper bound due to the red-noise--SGWB degeneracy in this small array configuration.}\label{fig:posterior_comparison}
\end{figure}

To quantify these effects in greater detail, Figure~\ref{fig:posterior_comparison} provides a more detailed view of the marginal posterior distributions, clearly showing the progressive improvements introduced by the hierarchical framework and the decorrelation procedure. 
For the red-noise parameters, the effect is pronounced: the baseline configuration with $\pi(\bth)$ (green) yields broad, weakly informative posteriors, for both $\log_{10}A_{\rm{RN}}$ and $\gamma_{\rm{RN}}$, which span nearly the entire prior range. The introduction of hierarchical priors $\pi(\bth|\bL)$ (orange) significantly tightens these distributions and shifts their peaks toward the injected values. 
The orthogonal reparametrization $\pi(\bth|\tilde\bL)$ (blue) further enhances both the precision and accuracy of the inference. A similar, though less pronounced, trend is observed for the DM variation parameters: while all three configurations yield reasonably constrained posteriors thanks to the additional chromatic information, 
the hierarchical treatment and decorrelation still produce a systematic shift toward the injected values, improving the accuracy without substantially altering the posterior widths. In contrast, the SGWB parameters remain essentially invariant across all three configurations. 
Both $\gamma_{\rm{GWB}}$ and $\log_{10}A_{\rm{GWB}}$ show nearly identical posteriors regardless of the noise modeling approach, 
with the amplitude remaining biased toward the upper prior bound. 
This confirms the conclusion drawn from Figure~\ref{fig:corner}: 
in this minimal 3-pulsar array, the hierarchical treatment successfully improves the inference of individual noise components 
but cannot fully overcome the fundamental red-noise--SGWB degeneracy that affects the common signal.

In Table~\ref{tab:summary_stats} we present the summary statistics of the posteriors across the three configurations, including median values, 68\% and 95\% credible intervals, and the Prior-to-Evidence (PTE) ratio. We use the PTE as a compact metric to assess the recovery of the injected fiducial values, quantifying the relative role of the likelihood with respect to the prior in shaping the posterior distribution. 
Values of the PTE close to 0.5 indicate that the posterior is jointly informed by prior and data, corresponding to an effective recovery of the injected value, while values approaching 0 or 1 signal posteriors dominated by either the prior or the likelihood, respectively.
The numerical results confirm and quantify the qualitative trends observed in Figures~\ref{fig:corner} and \ref{fig:posterior_comparison}. 
For the red-noise spectral index $\gamma_{\rm{RN}}$, the PTE increases from values $\simeq 0.43$ in the fixed-prior configuration to $\simeq 0.50$ after decorrelation, consistently with the progressive reduction of the 68\% credible interval from $^{+2.221}_{-1.844}$ to $^{+1.095}_{-1.422}$ and with the shift of the median toward the injected value $\gamma_{\rm{RN}}=3$. 
A similar behavior is observed for the red-noise amplitude $\log_{10}A_{\rm{RN}}$, whose PTE remains close to zero in all configurations, reflecting the limited informativeness of the data on this parameter in a three-pulsar setup, despite the improved concentration of the posterior and the partial recovery of the injected value.
For the DM variation parameters, the PTE exhibits a systematic increase across configurations, in particular for $\gamma_{\rm{DM}}$, where it rises from $\simeq 0.10$ to $\simeq 0.32$, consistently with the progressive shift of the median toward the injected value. 
In contrast, the SGWB parameters display PTE values very close to zero in all three frameworks, indicating that their posteriors remain strongly influenced by prior boundaries. 
This behavior quantitatively reflects the persistence of the red-noise--SGWB degeneracy in this minimal array, and confirms that, while the hierarchical treatment and decorrelation improve the recovery of individual noise components, they cannot overcome the intrinsic limitations affecting the inference of the common signal with only three pulsars.

\begin{table}[htbp]
\centering
\begin{tabular}{|c|c|c|c|c|c|}
\toprule
PARAMETER & SCENARIO & MEDIAN & 68\% & 95\% & PTE\\
\midrule
 & $\pi(\bth)$ & 2.684 & $^{+2.221}_{-1.844}$ & $^{+3.747}_{-2.552}$ & 0.434248\\
 \cline{2-5}
 & & & & \\[-8pt]
$\gamma_{\rm RN}$ & $\pi(\bth|\bL)$ & 2.714 & $^{+1.583}_{-1.351}$ & $^{+2.837}_{-2.031}$ & 0.442761\\
 \cline{2-5}
 & & & & \\[-8pt]
 & $\pi(\bth|\tilde\bL)$ & 3.008 & $^{+1.095}_{-1.422}$ & $^{+2.405}_{-2.768}$ &0.497350 \\
\midrule
 & $\pi(\bth)$ & $-14.859$ & $^{+0.973}_{-0.783}$ & $^{+1.548}_{-1.086}$ &0.007742\\
 \cline{2-5}
 & & & & \\[-8pt]
$\log_{10}A_{\rm{RN}}$ & $\pi(\bth|\bL)$ & $-14.227$ & $^{+0.594}_{-0.634}$ & $^{+0.939}_{-1.227}$ &0.006013\\
 \cline{2-5}
 & & & & \\[-8pt]
 & $\pi(\bth|\tilde\bL)$ & $-14.231$ & $^{+0.580}_{-0.671}$ & $^{+0.941}_{-1.322}$ & 0.007362\\
\midrule
 & $\pi(\bth)$ & 1.916 & $^{+0.299}_{-0.275}$ & $^{+0.616}_{-0.525}$ &0.102925\\
 \cline{2-5}
 & & & & \\[-8pt]
$\gamma_{\rm DM}$ & $\pi(\bth|\bL)$ & 1.974 & $^{+0.303}_{-0.269}$ & $^{+0.593}_{-0.532}$ &0.143399\\
 \cline{2-5}
 & & & & \\[-8pt]
 & $\pi(\bth|\tilde\bL)$ & 2.147 & $^{+0.326}_{-0.285}$ & $^{+0.601}_{-0.534}$ &0.315960\\
 \midrule
 & $\pi(\bth)$ & $-13.261$ & $^{+0.053}_{-0.050}$ & $^{+0.106}_{-0.097}$ &0.212537\\
 \cline{2-5}
 & & & & \\[-8pt]
$\log_{10}A_{\rm{DM}}$ & $\pi(\bth|\bL)$ & $-13.261$ & $^{+0.054}_{-0.050}$ & $^{+0.109}_{-0.100}$ &0.212157\\
 \cline{2-5}
 & & & & \\[-8pt]
 & $\pi(\bth|\tilde\bL)$ & $-13.314$ & $^{+0.055}_{-0.052}$ & $^{+0.107}_{-0.100}$ &0.408716\\
 \midrule
 & $\pi(\bth)$ & 3.268 & $^{+0.294}_{-0.262}$ & $^{+0.618}_{-0.500}$ &0.000887\\
 \cline{2-5}
 & & & & \\[-8pt]
$\gamma_{\rm SGWB}$ & $\pi(\bth|\bL)$ & 3.245 & $^{+0.238}_{-0.228}$ & $^{+0.506}_{-0.424}$ &0.001699\\
 \cline{2-5}
 & & & & \\[-8pt]
 & $\pi(\bth|\tilde\bL)$ & 3.194 & $^{+0.259}_{-0.258}$ & $^{+0.590}_{-0.478}$ & 0.000883\\
 \midrule
 & $\pi(\bth)$ & $-13.245$ & $^{+0.082}_{-0.095}$ & $^{+0.152}_{-0.204}$ &0.002402\\
 \cline{2-5}
 & & & & \\[-8pt]
$\log_{10}A_{\rm{SGWB}}$ & $\pi(\bth|\bL)$ & $-13.188$ & $^{+0.066}_{-0.075}$ & $^{+0.128}_{-0.161}$ & 0.000392\\
 \cline{2-5}
 & & & & \\[-8pt]
 & $\pi(\bth|\tilde\bL)$ & $-13.216$ & $^{+0.069}_{-0.090}$ & $^{+0.136}_{-0.215}$ & 0.001178\\
\bottomrule
\end{tabular}
\caption{Summary statistics for the noise and SGWB parameters of pulsar J1744$-$1134 across the three inference configurations: fixed uniform priors $\pi(\bth)$, hierarchical priors $\pi(\bth|\bL)$, and reparametrization $\pi(\bth|\tilde\bL)$. 
For each parameter, we report the posterior median, the symmetric 68\% and 95\% credible intervals, and the Prior-to-Evidence (PTE) ratio, used here as a diagnostic of the recovery of the injected fiducial values. 
PTE values close to 0.5 indicate posteriors jointly informed by prior and likelihood, while values close to 0 or 1 correspond to prior- or likelihood-dominated regimes, respectively. 
The progressive tightening of the red-noise posteriors and the systematic shift toward the injected values are reflected in increasing PTE values for the corresponding parameters, whereas the SGWB parameters exhibit consistently low PTE values, in agreement with the limited constraining power of this three-pulsar array on the common signal.}
\label{tab:summary_stats}
\end{table}

\begin{figure}[t]
    \centering
    \includegraphics[width=\textwidth]{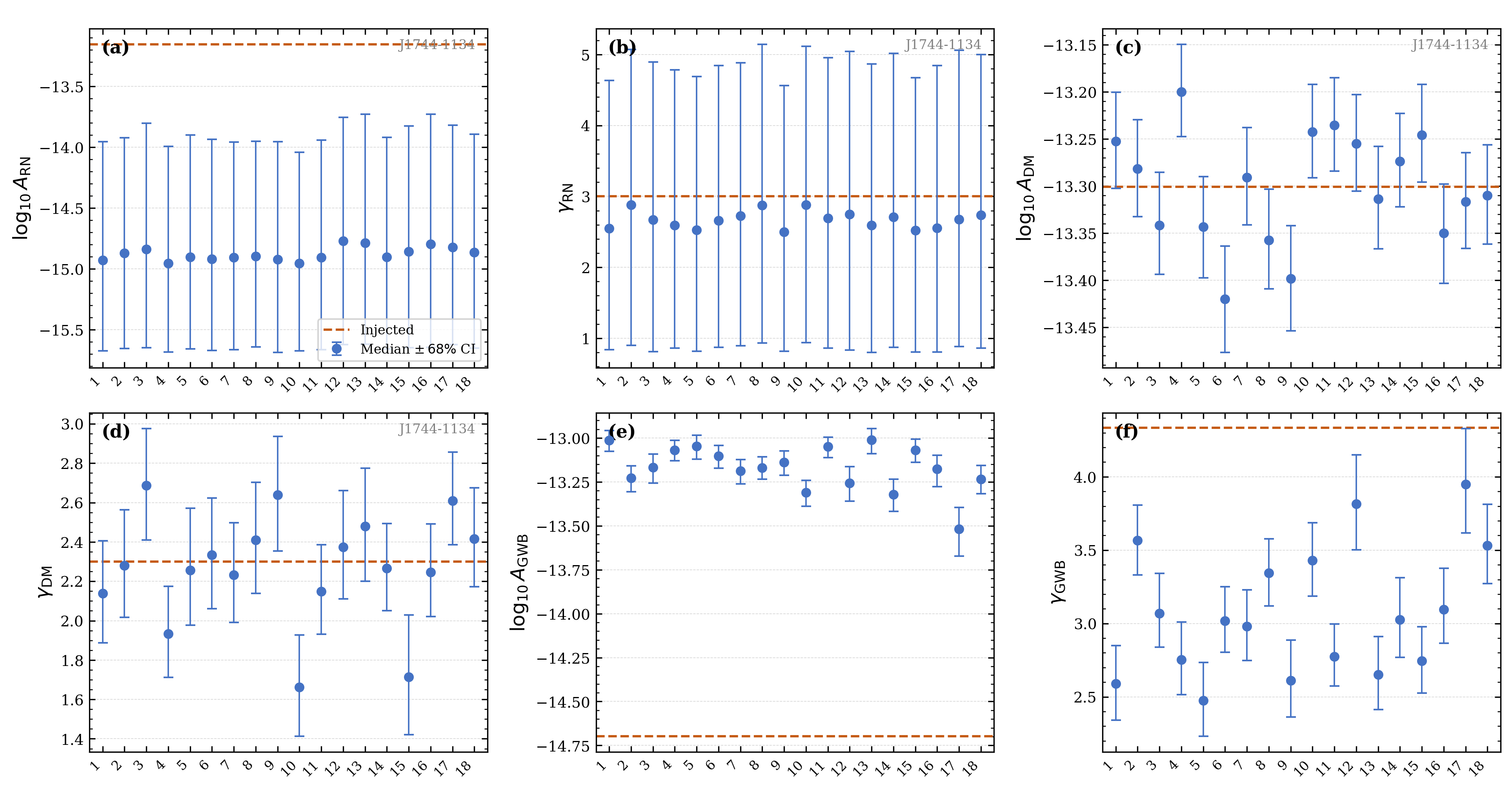}
    \caption{Parameter recovery across 18 independent realizations of the 
    baseline fixed-prior configuration $\pi(\bm{\vartheta})$, each generated 
    with a different random seed for the injection of the GWB, red noise, and 
    DM variation signals. Each point shows the posterior median with asymmetric 
    68\% credible intervals for pulsar J1744$-$1134 (panels a - d) and for the 
    common SGWB signal shared across the three-pulsar array (panels e and f). 
    Orange dashed lines mark the injected fiducial values. 
    The systematic bias in $\log_{10} A_{\rm GWB}$ and the specular offset 
    in $\log_{10} A_{\rm RN}$, both reproduced consistently across all 
    realizations, confirm that these features reflect the intrinsic 
    red-noise--SGWB degeneracy of the minimal three-pulsar configuration 
    rather than a statistical fluctuation of the single dataset.}
    \label{fig:recovery_gwb.png}
\end{figure}

Finally, to assess whether the systematic offsets observed in the posterior medians 
reflect a genuine structural feature of the inference or merely an artifact of a 
single unlucky realization, we performed 18 independent runs of the baseline 
fixed-prior configuration $\pi(\bm{\vartheta})$, each generated with a different 
random seed for the injection of the GWB, red noise, and DM variation signals. 
Figure~\ref{fig:recovery_gwb.png} shows the posterior median and 68\% credible interval 
for the six noise and SGWB parameters of pulsar J1744$-$1134 across all realizations. 
The results are remarkably consistent: the bias in $\log_{10} A_{\rm GWB}$ 
(panel~e) is reproduced across all runs, with the posterior median systematically displaced above the injected value and the injected value lying well outside the 68\% credible interval in every case. This behavior is consistent with the expected red-noise--SGWB degeneracy: in a minimal three-pulsar array, the HD correlation pattern provides insufficient leverage to disentangle the two contributions, and the sampler tends to allocate low-frequency power preferentially to the GWB. Correspondingly, the red-noise amplitude $\log_{10} A_{\mathrm{RN}}$ (panel a) displays the compensating offset, with all posterior medians systematically falling below the injected value, consistent with power being transferred from the red-noise sector to the GWB. The DM variation parameters (panels c and d) are instead reasonably recovered, with the injected values enclosed within the 68\% credible interval in the majority of runs, as expected from the fact that DM variations carry independent chromatic information and are not affected by the red-noise--SGWB degeneracy. The red-noise spectral index $\gamma_{\mathrm{RN}}$ (panel b) remains essentially prior-dominated across all realizations, with credible intervals spanning nearly the full prior range, while $\gamma_{\mathrm{GWB}}$ (panel f) shows larger run-to-run scatter but a consistent underestimation with respect to the injected value of $13/3$. Taken together, these results indicate that the biases showed in Figure~\ref{fig:posterior_comparison} are not a statistical fluctuation of the single dataset analyzed in this work, but a reproducible consequence of the intrinsic degeneracy structure and the limited constraining power of a three-pulsar array on the common signal.

%%%%%%%%%%%%%%%%%%%%%%%%%%%%%%%%%%%%%%%%%%%%%%%%%%%%%%%%%%%%%%%%
\section{Concluding remarks and future directions}\label{sec:conclusions}

In this work, we have developed a hierarchical Bayesian framework for PTA noise modeling that addresses the issue of prior sensitivity. 
The starting point is the introduction of hyperpriors on pulsar noise parameters, which elevates the problem to a hierarchical setting where hyperparameters themselves play the role of nuisance parameters. 
To handle the correlations between hyperparameters and physical noise parameters, we implemented an orthogonal reparameterization strategy based on a two-step Normalizing Flows algorithm. 
The algorithm learns the transformation that removes the projection of hyperparameters onto the subspace spanned by the physical parameters, thus producing a decorrelated set of hyperparameters as the orthogonal complement. 
This reparametrization is designed to weaken the coupling between physical parameters and hyperparameters, thereby mitigating the sensitivity of the marginal posterior of the physical parameters to the specific functional form of the noise hyperprior. We stress that reducing this coupling does not amount to removing the hierarchical information flow. The orthogonal projection acts at the level of the prior samples used to train the Normalizing Flows, and enforces sample-level decorrelation between physical parameters and transformed hyperparameters. Despite the orthogonal projection being singular and thus not invertible in closed form, the Normalizing Flows-based representation is invertible by construction and provides an exact Jacobian: the reparametrization thus preserves the full joint content of the hierarchical prior. As a consequence, shrinkage and inter-pulsar pooling are retained in the reparametrized representation, and continue to operate through the joint structure of the learned conditional prior across the stacked pulsar parameters. A detailed discussion of how prior sensitivity and population-level information coexist in our framework is provided in\ref{app:prior_sensitivity}.

Bayesian inference was performed using \texttt{i-nessai}, a flow-based nested sampling algorithm that, integrated in \texttt{Enterprise}, 
significantly accelerates the inference pipeline compared to traditional \texttt{PTMCMC} samplers while maintaining robust posterior and evidence estimation, as shown in \citet{nessai-spoke3}. 
We applied this framework to a minimal simulated dataset consisting of 3 pulsars, performing a simultaneous inference of noise and SGWB parameters. 
Although such a small array is not at all sufficient for a fully representative PTA analysis, it provides a controlled and transparent proof of concept that allows us to isolate and assess the key effects of the hierarchical modeling and reparametrization.

Our analyses - including corner plots, marginal posteriors, independence score metrics, credible intervals and PTE - consistently show the same picture. We find that: 
(i) introducing hyperpriors on the noise parameters significantly improves the inference of individual noise components, achieving both enhanced precision (reduced credible intervals) and improved accuracy (systematic shift of posterior medians toward injected values), with uncertainty reductions of up to 50\% for the red-noise parameters;
(ii) the orthogonal reparametrization via Normalizing Flows successfully decorrelates physical parameters from hyperparameters, as confirmed by independence scores approaching unity and, in this controlled setup, leads to further tightening the constraints;
(iii) both the hierarchical treatment and the decorrelation preserve the intrinsic $\log_{10}A$--$\gamma$ anticorrelation characteristic of power-law processes, validating that the reparametrization acts only on the hierarchical layer without disrupting genuine physical degeneracies encoded in the likelihood.
(iv) The SGWB parameters remain essentially unchanged across all configurations, with posteriors consistently biased toward the prior upper bound, confirming that the fundamental red-noise--SGWB degeneracy cannot be broken with such a limited array size.

In summary, this study represents a first application of our hierarchical and reparametrized framework to PTA data analysis. 
Although the present configuration serves primarily as a methodological test, the results clearly demonstrate the potential of the approach for reducing prior dependence and improving the interpretability of hierarchical PTA models. 
Future work will explore its application to larger arrays and to real PTA datasets, where the combined effects of hierarchy and reparametrization can be fully exploited to achieve more robust and physically meaningful inferences on the SGWB.

\section*{Acknowledgements}
\textit{This paper is supported by the Fondazione ICSC}, Spoke-3 Astrophysics and Cosmos Observations, \textit{National Recovery and Resilience Plan (Piano Nazionale di Ripresa e Resilienza, PNRR) Project ID CN\_00000013 ``Italian Research Center on High-Performance Computing, Big Data and Quantum Computing'' funded by MUR Missione 4 Componente 2 Investimento 1.4: Potenziamento strutture di ricerca e creazione di ``campioni nazionali di R\&S (M4C2-19 )'' - Next Generation EU (NGEU)}.

We acknowledge the Istituto Nazionale di Astrofisica (INAF) computing time allocation award under the the proposal \textit{FastParam} for the \textit{PLEIADI} computing infrastructure initiative. The HPC tests and benchmarks of this work have been carried out on the cluster \textit{PLEIADI}, installed and managed by INAF. The cluster we used is equipped with dual Intel Xeon E5-2697 V4 processors (36 cores per node) and 128~GB of RAM. For this work, all jobs were executed on a single node using 1 CPU core and 128~GB of RAM, via the SLURM scheduler within the \textit{large} partition.

The authors thank Romina T. Anfuso, Project Manager at Koexai srl, for project management support and coordination.

E.V. thanks Marco Bonici for insightful discussions about our reparametrization approach compared to the one in \citet{Paradiso_2025}, Sara Angela Filippini and Marco Lombardi for useful comments about the reparametrization obtained via orthogonal projection.

We are grateful to both the Referees for their careful reading and thoughtful comments. The revision process prompted by their reports led to a considerable improvement in the clarity and overall quality of the manuscript.

 \appendix

\section{Orthogonal projection algebra} \label{app:projection_algebra}
In this Appendix we provide the mathematical details of the orthogonal projection procedure described in Section~\ref{sec:orto}, 
explicitly specifying the relevant vector spaces and their dimensions, and prove that the transformed hyperparameters $\tL$ are orthogonal to the physical parameters $\bth$ by construction.
We consider $N$ independent samples of the $n$ physical parameters $\bth$ and of the $m$ hyperparameters $\bL$. The samples are organized as matrices:
\begin{equation}
\bth = 
\begin{pmatrix}
\vartheta_1^{(1)} & \vartheta_2^{(1)} & \cdots & \vartheta_n^{(1)} \\
\vartheta_1^{(2)} & \vartheta_2^{(2)} & \cdots & \vartheta_n^{(2)} \\
\vdots & \vdots & \ddots & \vdots \\
\vartheta_1^{(N)} & \vartheta_2^{(N)} & \cdots & \vartheta_n^{(N)}
\end{pmatrix} \in \mathbb{R}^{N \times n},
\quad
\bL = 
\begin{pmatrix}
\Lambda_1^{(1)} & \Lambda_2^{(1)} & \cdots & \Lambda_m^{(1)} \\
\Lambda_1^{(2)} & \Lambda_2^{(2)} & \cdots & \Lambda_m^{(2)} \\
\vdots & \vdots & \ddots & \vdots \\
\Lambda_1^{(N)} & \Lambda_2^{(N)} & \cdots & \Lambda_m^{(N)}
\end{pmatrix} \in \mathbb{R}^{N \times m},
\end{equation}
where each row represents one single draw of all the parameters, and each column represents $N$ draws of a single parameter. 

The orthogonal projection operates in the $N$-dimensional draw space $\mathbb{R}^N$, 
projecting vectors onto the subspace $\mathcal{S}_{\bth}$ spanned by the physical 
parameter draws. The projector operator is defined as:
\begin{equation}
P_{\bth} = \bth \left( \bth^T \bth \right)^{-1} \bth^T \in \mathbb{R}^{N \times N},
\end{equation}
which is the standard orthogonal projector onto the subspace $\mathcal{S}_{\bth} = 
\text{span}\{\bth_1, \bth_2, \ldots, \bth_n\} \subset \mathbb{R}^N$, where $\bth_j$ 
denotes the $j$-th column of $\bth$. The transformed hyperparameters are then obtained as:
\begin{equation}
\tL = \bL - P_{\bth}\bL= \left( I_N - P_{\bth} \right) \bL \in \mathbb{R}^{N \times m},
\label{eq:projection_matrix}
\end{equation}
where $I_N$ is the $N \times N$ identity matrix. Geometrically, this operation acts independently on each of the $m$ columns of $\bL$: 
for each hyperparameter $\bL_k$ (the $k$-th column of $\bL$), the projector removes 
the component lying in $\mathcal{S}_{\bth}$ and retains only the component in the 
orthogonal complement $\mathcal{S}_{\bth}^{\perp}$. Thus, each column of $\tL$ lies 
entirely in the orthogonal complement $\mathcal{S}_{\bth}^{\perp} \subset \mathbb{R}^N$.

We now demonstrate that the transformed hyperparameters $\tL$ are orthogonal to the 
physical parameters $\bth$. The orthogonality condition is:
\begin{equation}
\bth^T \tL = \mathbf{0}_{n \times m},
\label{eq:orthogonality_condition}
\end{equation}
where:
\begin{itemize}
\item $\bth^T \in \mathbb{R}^{n \times N}$ (transpose of the $N \times n$ matrix $\bth$)
\item $\tL \in \mathbb{R}^{N \times m}$ (the transformed hyperparameters)
\item $\bth^T \tL \in \mathbb{R}^{n \times m}$ (matrix product)
\end{itemize}

By substituting $\tL$ from equation \eqref{eq:projection_matrix} and expanding the products, we have
\begin{align}
\bth^T \tL &= \bth^T \left( I_N - P_{\bth} \right) \bL \\
&= \bth^T \bL - \bth^T P_{\bth} \bL \\
&= \bth^T \bL - \bth^T \left[ \bth \left( \bth^T \bth \right)^{-1} \bth^T \right] \bL \\
&= \bth^T \bL - \left[ \bth^T \bth \left( \bth^T \bth \right)^{-1} \right] \bth^T \bL \\
&= \bth^T \bL - I_n \, \bth^T \bL \\
&= \bth^T \bL - \bth^T \bL \\
&= \mathbf{0}_{n \times m}.
\end{align}
This result can also be derived using the fundamental properties of the projection operator 
$P_{\bth}$, namely its idempotency ($P_{\bth}^2 = P_{\bth}$) and symmetry ($P_{\bth}^T = P_{\bth}$).

%%%%%%%%%%%%%%%%%%%%%%%%%%%%%%%%%%%%%%%%%%%%%%%%%%%%%%%%%%%%%%%%%%

\section{Prior sensitivity and orthogonal reparametrization}
\label{app:prior_sensitivity}
In this appendix we clarify in what sense the orthogonal reparametrization is expected to reduce the sensitivity of the posterior inference to the chosen hyperprior, without implying the removal of the hierarchical dependence itself. 

Note that in the following, $\bth \in \mathbb{R}^n$ and $\tL \in\mathbb{R}^m$ denote a single draw of the physical parameters and transformed hyperparameters, and $\pi(\bth|\tL)$ is a density in $\mathbb{R}^n$ conditional on $\tL\in\mathbb{R}^m$. This is distinct from the level at which the orthogonality condition of \ref{app:projection_algebra} is formulated: $\bth \in 
\mathbb{R}^{N \times n}$ and $\tL \in \mathbb{R}^{N \times m}$ collect $N$ 
independent draws of the physical parameters and transformed hyperparameters, 
respectively, and the orthogonality condition $\bth^T \tL = \mathbf{0}_{n 
\times m}$ is a statement about the $n \times 
m$ matrix of sample inner products computed on the training set of the Normalizing Flows. The relation between the two levels, i.e. sample orthogonality at the N-draw level versus functional dependence of the conditional density at the single-draw level, is discussed below, at the end of~\ref{subapp:gradient}.
We begin by reporting the joint posterior of the physical parameters $\bth$ and the transformed hyperparameters $\tL$ in the reparametrized hierarchical model 
\begin{equation}
\mathcal{P}(\bth,\tL | \bdt)
\propto
L(\bdt | \bth)\,
\pi(\bth | \tL)\,
\pi'(\tL),
\label{eq:joint_posterior_tL}
\end{equation}
where the likelihood $L(\bdt | \bth)$ depends only on the physical parameters $\bth$ and not on the hyperparameters $\tL$. The marginal posterior of $\bth$ is therefore
\begin{equation}
p(\bth | \bdt)
\propto
L(\bdt | \bth)
\int
\pi(\bth | \tL)\,\pi'(\tL)\,d\tL
\equiv
L(\bdt | \bth)\,\pi_m(\bth),
\label{eq:marginal_posterior_theta}
\end{equation}
where we define the induced marginal prior
\begin{equation}
\pi_m(\bth)
\equiv
\int
\pi(\bth | \tL)\,\pi'(\tL)\,d\tL.
\label{eq:marginal_prior_theta}
\end{equation}
Equation~\eqref{eq:marginal_posterior_theta} makes explicit that, once the likelihood is fixed, the dependence of the marginal posterior $p(\bth | \bdt)$ on the hyperprior $\pi'(\tL)$ is entirely mediated by the induced marginal prior $\pi_m(\bth)$.

We now consider a perturbation of the hyperprior,
\begin{equation}
\pi'(\tL) \;\longrightarrow\; \pi'(\tL) + \delta \pi'(\tL),
\qquad
\int \delta \pi'(\tL)\,d\tL = 0,
\label{eq:hyperprior_perturbation}
\end{equation}
where the second condition enforces normalization to first order. The corresponding first-order variation of the marginal prior is
\begin{equation}
\delta \pi_m(\bth)
=
\int
\pi(\bth | \tL)\,\delta \pi'(\tL)\,d\tL.
\label{eq:variation_marginal_prior}
\end{equation}
Equation~\eqref{eq:variation_marginal_prior} is exact and does not yet use the orthogonality condition $\bth^{\top}\tL = 0$. It shows that prior sensitivity is controlled entirely by how strongly the conditional prior $\pi(\bth | \tL)$ varies as a function of $\tL$. The role of the orthogonal reparametrization enters only at the interpretative level: it provides a geometric motivation for expecting a weaker dependence of $\pi(\bth | \tL)$ on $\tL$ in the transformed coordinates than in the original parametrization. This is precisely the sense in which our method addresses and is designed to mitigate prior dependence in the hierarchical model.

\subsection{Local prior insensitivity: gradient formulation}
\label{subapp:gradient}
Let $\tL_0$ be a reference point in hyperparameter space, for instance the mean of $\pi'(\tL)$. Assuming that $\pi(\bth | \tL)$ is differentiable in a neighborhood of $\tL_0$, we expand
\begin{equation}
\pi(\bth | \tL)
=
\pi(\bth | \tL_0)
+
\nabla_{\tL}\pi(\bth | \tL)\big|_{\tL_0}
\cdot
(\tL - \tL_0)
+
O\!\left(\|\tL-\tL_0\|^2\right).
\label{eq:taylor_conditional_prior}
\end{equation}
Substituting Eq.~\eqref{eq:taylor_conditional_prior} into Eq.~\eqref{eq:variation_marginal_prior}, we obtain
\begin{align}
\delta \pi_m(\bth)
&=
\pi(\bth | \tL_0)
\int \delta \pi'(\tL)\,d\tL
+
\nabla_{\tL}\pi(\bth | \tL)\big|_{\tL_0}
\cdot
\int (\tL-\tL_0)\,\delta \pi'(\tL)\,d\tL
\nonumber\\
&\hspace{1.5cm}
+
O\!\left(\|\tL-\tL_0\|^2\right).
\label{eq:variation_marginal_prior_expanded}
\end{align}
The zeroth-order term vanishes by the normalization condition in Eq.~\eqref{eq:hyperprior_perturbation}. Therefore, if the transformed conditional prior satisfies the \emph{local prior insensitivity} condition around $\tL_0$,
\begin{equation}
\nabla_{\tL}\pi(\bth | \tL)\big|_{\tL_0} = 0,
\label{eq:local_prior_insensitivity}
\end{equation}
then the first-order contribution also vanishes and
\begin{equation}
\delta \pi_m(\bth)
=
O\!\left(\|\tL-\tL_0\|^2\right).
\label{eq:second_order_variation_marginal_prior}
\end{equation}

Under the same assumption, the variation of the normalized marginal posterior $p(\bth | \bdt)$ is also second order. Indeed, since
\begin{equation}
p(\bth | \bdt)
=
\frac{L(\bdt | \bth)\,\pi_m(\bth)}
{\int L(\bdt | \bth)\,\pi_m(\bth)\,d\bth},
\label{eq:normalized_marginal_posterior}
\end{equation}
a second-order variation of $\pi_m(\bth)$ induces a second-order variation of both numerator and normalization constant, and therefore of the posterior itself.

The importance of Eq.~\eqref{eq:local_prior_insensitivity} is conceptual: it identifies a sufficient local condition under which perturbations of the hyperprior have a suppressed effect on the marginal inference of $\bth$. The orthogonality relation $\bth^{\top}\tL=0$ is not used here as a direct algebraic input to derive Eq.~\eqref{eq:local_prior_insensitivity}; rather, it motivates the expectation that the transformed coordinates define a weaker-coupling representation, in which local prior insensitivity is more plausible than in the original parametrization. The connection between the two levels can be made explicit as follows. The condition $\bth^{\top}\tL = 0$ holds in the $N$-dimensional draw space $\mathbb{R}^N$, as demonstrated in Appendix~\ref{app:projection_algebra}: for any finite set of $N$ draws, the sample inner product between $\bth$ and $\tL$ is exactly zero by construction. Zero sample correlation is not equivalent to, but is consistent with, a weak functional dependence of $\pi(\bth | \tL)$ on $\tL$: if the draws of $\bth$ carry no linear information about $\tL$, it is natural to expect that the conditional prior $\pi(\bth | \tL)$ varies weakly as $\tL$ is perturbed around its mean. In our framework, this transformed conditional prior is represented explicitly by a dedicated conditional Normalizing Flow, which learns the residual non-linear dependence of $\bth$ on $\tL$ that survives after the orthogonal projection.

\subsection{Local prior insensitivity: information-based formulation}

A complementary way to characterize weak local coupling between $\bth$ and $\tL$ is through the cross-information matrix of the conditional prior, given by
\begin{equation}
I_{\bth\tL}^{(\pi)}
=
\mathbb{E}_{\pi(\bth,\tL)}
\left[
\frac{\partial \log \pi(\bth | \tL)}{\partial \bth}
\frac{\partial \log \pi(\bth | \tL)}{\partial \tL^{\top}}
\right]\,,
\label{eq:prior_cross_information}
\end{equation}
where the expectation is taken with respect to the joint prior
$\pi(\bth,\tL)=\pi(\bth | \tL)\,\pi'(\tL)$. Equation~\ref{eq:prior_cross_information} therefore quantifies the cross-information between the score directions associated with the physical parameters and the transformed hyperparameters under the hierarchical prior: a small or vanishing $I_{\bth\tL}^{(\pi)}$ means that the score directions associated with $\bth$ and $\tL$ are weakly coupled under the joint prior. This condition is analogous in spirit to the orthogonality criterion of \citet{cox-reid-1987}, although here it is applied to the prior layer of the hierarchical model rather than to the likelihood. As in the gradient formulation, Eq.~\eqref{eq:prior_cross_information} is not derived directly from the sample-space relation $\bth^{\top}\tL=0$; rather, it provides an information-based characterization of the same weak-coupling regime that the orthogonal reparametrization is designed to promote.

In summary, the exact identity in Eq.~\eqref{eq:variation_marginal_prior} shows that prior sensitivity is controlled by the $\tL$-dependence of the conditional prior $\pi(\bth | \tL)$. The orthogonal reparametrization is designed to promote a weak-coupling regime in which $\pi(\bth | \tL)$ is expected to satisfy local prior insensitivity more closely than in the original parametrization, so that perturbations of the hyperprior have a reduced impact on the marginal posterior of $\bth$. 

\subsection*{}
We conclude with two remarks on how population-level information is preserved and on the interplay between local prior insensitivity and hierarchical pooling. First, the orthogonal transformation is implemented through invertible 
Normalizing Flows with tractable Jacobian. As a change of variables from 
$(\bth,\bL)$ to $(\bth,\tL)$, it preserves by construction the full joint 
content of the hierarchical prior $\pi(\bth|\bL)\,\pi'(\bL)$, which is 
simply re-expressed in the new coordinates as $\pi(\bth|\tL)\,\tilde\pi'
(\tL)$. In this strict sense, no information encoded in the original 
hierarchical structure is lost: the learned representation is 
mathematically equivalent to the original one, up to the reparametrization. Second, and more substantively, shrinkage and inter-pulsar pooling operate 
through the joint structure of $\pi(\bth|\tL)$ across the stacked pulsar 
parameters. We recall that $\bth = \{\gamma_{\mathrm{RN}}, 
\gamma_{\mathrm{DM}}, \log_{10}A_{\mathrm{RN}}, \log_{10}A_{\mathrm{DM}}\}$ 
collects the noise parameters of all pulsars in the array, so that $\bth 
\in \mathbb{R}^n$ with $n = n_p \times 4$. Hierarchical pooling manifests 
itself in the correlations that $\pi(\bth|\tL)$ induces \emph{among the 
components of $\bth$ corresponding to different pulsars}, rather than 
solely in a strong functional dependence of $\pi(\bth|\tL)$ on $\tL$. 
Local prior insensitivity and residual inter-pulsar coupling are therefore 
not in contradiction: they act along different directions of the joint 
structure of $\pi(\bth|\tL)$, and the non-linear conditional Normalizing 
Flow is in principle expressive enough to accommodate both. Shrinkage is 
preserved to the extent that this inter-pulsar correlation structure 
survives in the learned representation. The balance between weak $\tL$-dependence and preservation of inter-pulsar pooling in the reparametrized model is ultimately a quantitative question, whose full assessment lies beyond the scope of the present proof-of-concept study and will be addressed in future work on larger arrays, where prior 
sensitivity and population-level information pooling can be explored in a more informative regime.

%%%%%%%%%%%%%%%%%%%%%%%%%%%%%%%%%%%%%%%%%%%%%%%%%%%%%%%
 \section{\texttt{I-nessai} diagnostic for hierarchical Bayesian inference} \label{app:state_plot}

In this appendix we present the evolution of key diagnostic quantities throughout the sampling process displayed in the state plots produced by \texttt{i-nessai} for the runs with hierarchical priors before and after orthogonal reparametrization, configurations $\pi(\bth|\bL)$ and $\pi(\bth|\tL)$ of Section~\ref{sec:results} respectively. Both runs employed importance nested sampling with Normalizing Flows using a RealNVP architecture consisting of 6 coupling blocks, 4 layers per block, and 64 neurons per layer.

\subsection[Configuration with hierarchical priors]{Configuration with hierarchical priors $\pi(\bth|\bL)$}

Figure~\ref{fig:state8} shows the state plot for the hierarchical prior run with $N_{\rm live} = 16000$ live points. The top panel displays the log-likelihood evolution, with the minimum log-likelihood (blue solid line) increasing with some fluctuations in the early iterations from approximately $48000$ to the plateau around $\log \mathcal{L} \approx 52500$, indicating successful exploration of the parameter space. The second panel shows likelihood constraint, which increases monotonically as expected during the nested sampling process.

The third panel presents the log-evidence ($\log Z$) and its uncertainty ($\log dZ$). The evidence converges smoothly to $\log Z \approx 53060$, with the uncertainty decreasing from $\sim 1$ to $\sim 10^{-2}$. The fourth panel shows that the run required approximately $7.5 \times 10^5$ likelihood evaluations, completed in about 3 hours and 16 minutes. For reference, our baseline configuration with fixed priors is completed in about 2 hours and 37 minutes for around $10^6$ likelihood evaluations.

The fifth panel displays the Effective Sample Size (ESS). The live points ESS (red dashed) decreases from the initial value of $\sim 16000$ as sampling progresses, while the posterior ESS (blue solid) remains low initially but increases to approximately $7000$-$8000$ in the final iterations, providing a reasonable number of effective posterior samples. The sixth panel shows the level importance, with both total and posterior evidence contributions increasing significantly in the latter half of the run, as expected.

The seventh panel confirms that the number of live points (red dashed) remains constant at 16000 throughout the run, with removed (blue) and added (orange) samples tracking this value. The bottom panel shows the stopping criterion ratio, which peaks around iteration 20 before decreasing to zero, confirming proper termination.

\subsection[Configuration with hierarchical prior after decorrelation]{Configuration with hierarchical prior $\pi(\bth|\tL)$ after decorrelation}

\begin{figure}[htbp]%%
\hspace{-1.5cm}
\includegraphics[width=1.2\textwidth]{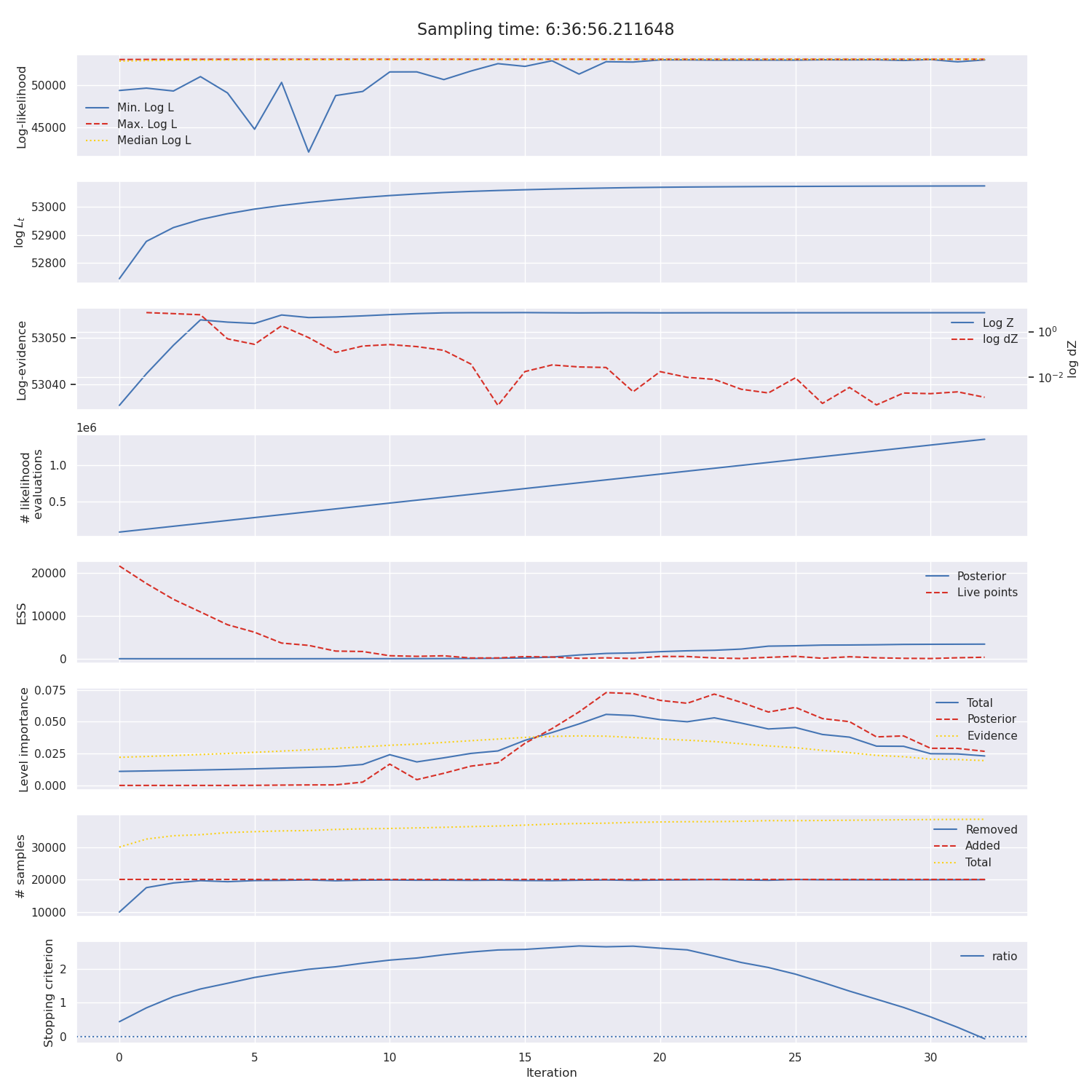}
%% Use \caption command for figure caption and label.
\caption{State plot from \texttt{i-nessai} for the simulated dataset comprising 3 pulsars with hierarchical priors $\pi(\bth|\tL)$ reparametrized according to the orthogonal projection.}\label{fig:state9}
\end{figure}

Figure~\ref{fig:state9} shows the state plot for the reparametrized run with $N_{\rm live} = 20000$ live points. The run required approximately twice the computational time (6 hours and 37 minutes) compared to the previous configuration , despite using only 25\% more live points. However, the sampling exhibits notably regular behavior.

\begin{figure}[htbp]
\hspace{-1.5cm}
\includegraphics[width=1.2\textwidth]{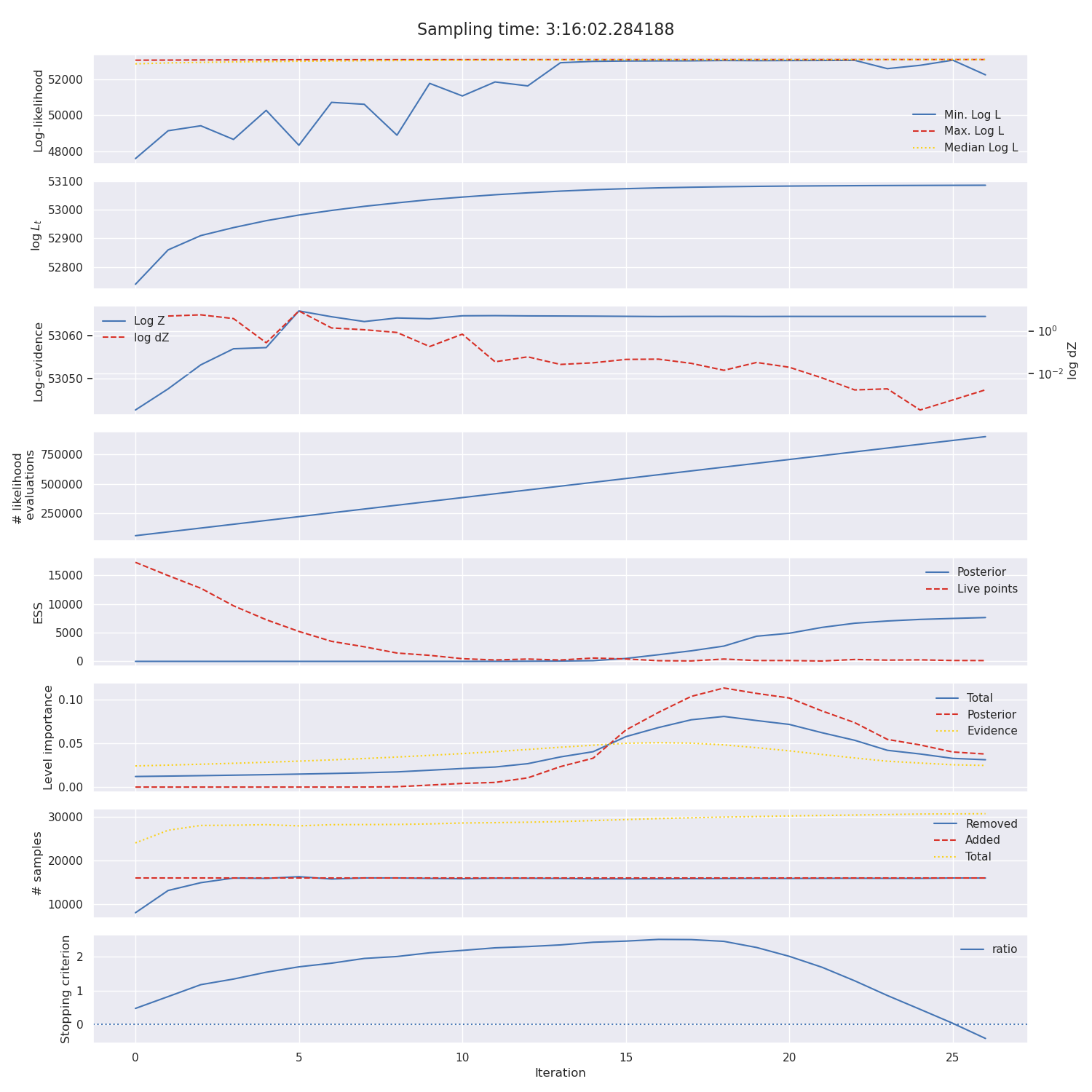}
%% Use \caption command for figure caption and label.
\caption{State plot from \texttt{i-nessai} for the simulated dataset comprising 3 pulsars with hierarchical priors $\pi(\bth|\bL)$ before the decorrelation.}\label{fig:state8}
\end{figure}

The top panel shows a smoother evolution of the minimum log-likelihood, which increases steadily from approximately $50000$ to plateau around $\log \mathcal{L} \approx 53000$, with significantly reduced fluctuations compared to Figure~\ref{fig:state8}. The second panel shows the same monotonic increase in $\log L$.

The evidence converges smoothly to $\log Z \approx 53060$, consistent with the third panel of Figure~\ref{fig:state9}, with the uncertainty again decreasing to $\sim 10^{-2}$. The run required approximately $1.2 \times 10^6$ likelihood evaluations (fourth panel), reflecting the increased computational cost.

The fifth panel shows the ESS behavior. The posterior ESS remains substantially lower throughout most of the run, increasing only in the final iterations. This reflects the exploration of transformed parameter spaces, where the Normalizing Flow must learn the complex geometry of the reparametrized posterior. To compensate for this lower ESS, we increased the number of live points to 20000 and performed careful posterior reweighting to ensure reliable parameter constraints despite the reduced effective sample count.

The level importance (sixth panel) shows a regular behavior, with contributions increasing toward the end of the run. The sample tracking (seventh panel) confirms stable operation with 20000 live points, and the stopping criterion (bottom panel) properly reaches zero after peaking around iteration 20.

Overall, both state plots demonstrate successful convergence of the nested sampling runs.

\section*{Declaration of generative AI and AI-assisted technologies in the manuscript preparation process}
During the preparation of this work, the authors used chatGPT5 to correct grammar misspellings, consistency of notation and terminology, and language fluency. After using this tool/service, the authors reviewed and edited the content as needed and assume full responsibility for the content of the published article.

\bibliographystyle{elsarticle-harv}
\bibliography{Manuscript}

\end{document}